\documentclass[aps,prl,preprint,groupedaddress]{revtex4-2}
\usepackage{graphicx}
\usepackage{amsmath}
\usepackage{mathtools}

\usepackage{xcolor}

\begin{document}

%\preprint{}

\title{Spatiotemporal steering of non-diffracting wavepackets}
% alternative: Spatiotemporal maneuvering of light bullets


\author{Haiwen Wang}
\email{hwwang@stanford.edu}
%\thanks{}
\affiliation{Department of Applied Physics, Stanford University, Stanford, CA, 94305, USA}

\author{Cheng Guo}
\affiliation{Ginzton Laboratory and Department of Electrical Engineering, Stanford University, Stanford, CA, 94305, USA}

\author{Shanhui Fan}
\email{shanhui@stanford.edu}
\affiliation{Ginzton Laboratory and Department of Electrical Engineering, Stanford University, Stanford, CA, 94305, USA}


\date{\today}

\begin{abstract}
We study the dynamics of space-time non-diffracting wavepackets, commonly known as light bullets, in a spatiotemporally varying medium. We show that by spatiotemporal refraction, a monochromatic focused beam can be converted to a light bullet that propagates at a given velocity. By further designing the index profile of the spatiotemporal boundary, the group velocity and the propagation direction of the light bullet can be engineered in a programmable way. All effects mentioned above cannot be achieved by spatial or temporal boundaries, and are only possible with spatiotemporal boundaries. These findings provide unique ways to engineer the dynamics of electromagnetic wavepackets in space-time. Such wavepackets with engineered spacetime trajectory may find potential applications in the spatiotemporal control of material properties or particles, or for use as a way to emulate relativistic physics in the laboratory.
\end{abstract}

%\keywords{}

\maketitle

\section{}
%\subsection{}
%\subsubsection{}

Controlling the propagation of electromagnetic waves is of great interest for many applications \cite{pendry2000negative, efremidis2019airy, dorrah2023light}. Traditionally, this has been achieved by designing spatial structures, ranging from macroscopic optical elements to microscopically engineered metamaterials \cite{JoannopoulosJohnsonWinnMeade+2008, yu2014flat, shaltout2019spatiotemporal, wang2020compact, wang2021engineering, wang2022design}. The temporal dimension provides another degree of freedom for controlling electromagnetic waves \cite{morgenthaler1958velocity, mendoncca2003temporal, yanik2004stopping, yanik2004time, xiao2014reflection, plansinis2015temporal, yuan2018synthetic, pacheco2020temporal, pacheco2020antireflection, dutt2020single, long2023time, tirole2023double}. Time-varying media can exhibit novel phenomena such as photonic time crystal \cite{lustig2018topological, lyubarov2022amplified}, photonic refrigeration \cite{buddhiraju2020photonic}, adiabatic frequency transfer \cite{yanik2004stopping, preble2007changing, tanabe2009dynamic, kampfrath2010ultrafast, howard2019photon, shcherbakov2019photon, zhou2020broadband}, temporal reflection and time reversal \cite{bacot2016time, vezzoli2018optical, moussa2023observation}, and the dynamic Casimir effect \cite{wilson2011observation}. Combining spatial and temporal structuring leads to spatiotemporally varying media \cite{hashimshony2001conversion, philbin2008fiber, biancalana2007dynamics, shi2016dynamic, subkhangulov2016terahertz, guo2019nonreciprocal, sharabi2022spatiotemporal, fan2023ultrafast}. In this case, the material property varies with both space and time, ultimately forming space-time metamaterials \cite{biancalana2007dynamics, caloz2019spacetime}. Such spatiotemporal media have enabled novel effects like terahertz wave generation \cite{hashimshony2001conversion}, optical analog of event horizon \cite{philbin2008fiber}, optical non-reciprocity \cite{guo2019nonreciprocal}, and ultrafast wavefront shaping \cite{fan2023ultrafast}.

The propagation dynamics of electromagnetic waves can also be controlled by designing the field structure itself. Recently, there are emerging interests in a type of electromagnetic wave with unique propagating dynamics, known as the space-time non-diffracting wavepacket, or simply as the light bullet \cite{lu1992nondiffracting, saari1997evidence, zamboni2002new, saari2004generation, hernandez2008localized, hernandez2013non, turunen2010propagation, kondakci2017diffraction, li2020velocity, li2020optical, bhaduri2020anomalous, guo2021structured, guo2021generation, yessenov2022space, pang2022synthesis, chen2022time, yessenov2023relativistic, yessenov2024experimental}. The group velocity $v_g$ of the light bullet can be either subluminal or superluminal \cite{kondakci2017diffraction, bhaduri2020anomalous, yessenov2024experimental}, and its pulse shape remains unchanged during propagation. Given its unique property, there has been great interest in generating the light bullet \cite{kondakci2017diffraction, guo2021structured}, controlling its dynamics \cite{yessenov2024experimental}, and studying how it interacts with various media \cite{bhaduri2020anomalous, chen2022time}. Most of the previous works focus on its interaction with a spatially structured medium. However, since the light bullet possesses a unique spatiotemporal correlation in its field, it should be interesting to study the interaction of a light bullet with a spatiotemporally structured medium.

In this work, we investigate the propagation dynamics of light bullets in spatiotemporally varying media. We point out that for light bullets, spatiotemporal boundaries generate novel effects that are not possible with spatial or temporal boundaries. These include launching light bullets from monochromatic beams, immobilizing propagating light bullets, and programmable back-and-forth motion. Controlling light bullets with spatiotemporal boundaries opens new possibilities for generating novel spatiotemporal light and engineering the space-time dynamics of electromagnetic fields. Our findings may find applications in controlling spatiotemporal properties of materials or particles, and in emulating relativistic physics in the laboratory.

\begin{figure}
\includegraphics[width=0.68\textwidth]{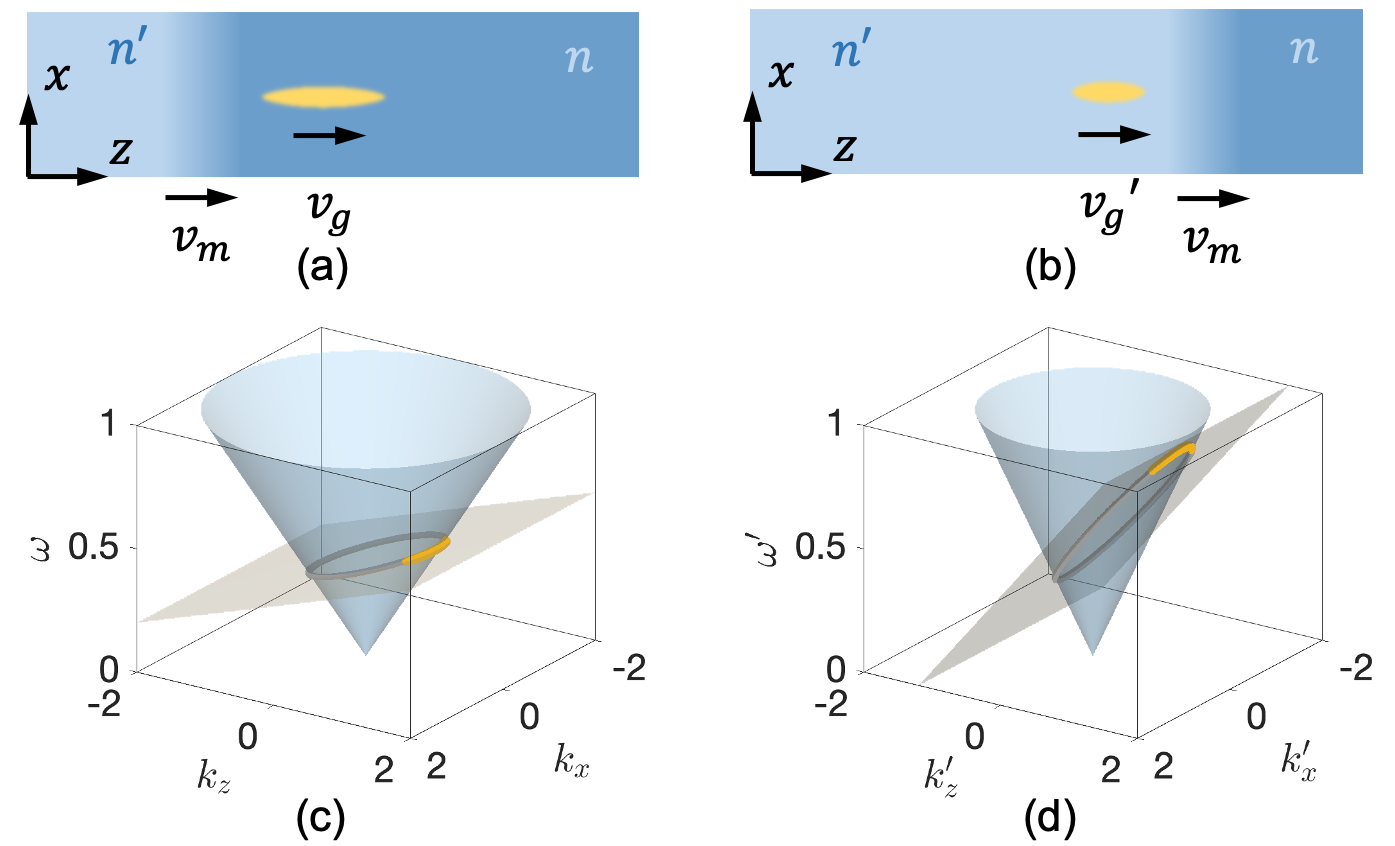}%
\caption{\label{fig_schematic} Schematic of a light bullet interacting with a spatiotemporal graded index boundary. (a) A light bullet (yellow, with velocity $v_g$) and a moving index front (blue background, with velocity $v_m$) before interaction. $n$ and $n'$ are the refractive indices before and after the boundary, respectively. (b) The light bullet (with velocity $v_g'$) and the moving index front after the interaction. In the schematic, we have assumed $v_m> v_g$ and $v_m> v_g'$. Our theory also applies to other relations between $v_m$, $v_g$, and $v_g'$.  (c) The spatiotemporal correlation of the light bullet before interaction. The light cone (blue) has an index of $n$. The plane wave components of the light bullet are shown in the continuum of points in yellow, which lie in a plane (gray) showing the spatiotemporal correlation. (d) Same as (c) but for the light bullet after interaction. The light cone (blue) has an index of $n'$.}
\end{figure}

A schematic of the process under study is shown in Figs.~\ref{fig_schematic}a and \ref{fig_schematic}b. The spatiotemporal boundaries we study here are index fronts that move in real space at some constant velocity $v_m$. Such boundaries are generalizations of the spatial or temporal boundaries since these two boundaries correspond to the limiting case of $v_m\rightarrow 0$ or $\infty$, respectively. In the following, the terms "spatiotemporal boundary" and "moving index front" are used interchangeably. We assume that at a given spatial point, as the front moves past, the refractive index changes from $n$ to $n'$. Consider a light bullet initially propagates in the homogeneous medium of index $n$, at a group velocity of $v_g$, with $v_g\neq v_m$. For simplicity, here we assume the velocities of both objects are along the $z$-axis, and we only consider one transverse dimension denoted as $x$. We also assume that the index change is adiabatic on a spatial scale larger than $\lambda_0$ or temporally larger than $\lambda_0/c$ to minimize spatiotemporal reflection. Such scale should be much smaller than the coherence length or pulse length of the light source in a possible experimental demonstration.

A light bullet in a homogeneous medium with index $n$ can be decomposed into a superposition of plane wave components. These plane wave components satisfy the following two relations. One is the dispersion relation (or the light cone, shown in blue in Fig.~\ref{fig_schematic}c) of the medium, given by:
\begin{equation}\label{lightcone}
    k_x^2+k_z^2=\left(\frac{n\omega}{c}\right)^2
\end{equation}
Here, $k_x$ and $k_z$ are wavevectors in $x$ and $z$ directions, respectively. $\omega$ is the angular frequency of the plane wave component. $c$ is the speed of light in vacuum. The second relation is the spatiotemporal correlation (shown in gray in Fig.~\ref{fig_schematic}c) \cite{kondakci2017diffraction, guo2021structured, yessenov2022space}, given by:
\begin{equation}\label{stcorrelation}
    \omega=v_g(k_z-k_0)+\omega_0
\end{equation}
Here, $v_g$ is the group velocity of the light bullet, $k_0$ and $\omega_0$ describe a constant offset in wavevector and angular frequency, respectively. The plane wave components of a light bullet before the interaction, satisfying both relations, are represented by the yellow points in Fig.~\ref{fig_schematic}c.

The behavior of each plane wave component under a spatiotemporal boundary is well-understood as a front-induced transition \cite{gaafar2019front}. A plane wave component represented by ($\omega, k_x, k_z$) on the light cone of index $n$ transits into a new plane wave component represented by ($\omega', k_x', k_z'$) on the light cone of index $n'$. During the transition, the transverse wavevectors are conserved, i.e. $k_x=k_x'$. The angular frequencies $\omega$, $\omega'$ and the longitudinal wavevectors $k_z$, $k_z'$ satisfies \cite{gaafar2019front}:
\begin{equation}\label{omegakz_conservation}
    \omega - v_m k_z = \omega' - v_m k_z'
\end{equation}

When a light bullet interacts with a moving index front, its plane wave components go through front-induced transitions independently, and the final state is represented by points on the light cone of index $n'$ (Fig.~\ref{fig_schematic}d). If we assume the wavepacket is paraxial ($k_x \ll k_z$, $k_x \ll k_z'$), the spatiotemporal correlation after the interaction with the moving index front is \cite{suppmat}:
\begin{equation}\label{stcorrelation2}
    \omega'=v_g'(k_z'-k_0')+\omega_0'
\end{equation}
$k_0'$, $\omega_0'$ represent a constant offset in wavevector and angular frequency after the interaction. $v_g'$ represents the group velocity of the light bullet after interaction and is given by:
\begin{equation}
    v_g'(v_g, n, n', v_m) = \frac{\frac{c}{n}(v_m-v_g)(v_m-\frac{c}{n})+v_m(v_m-\frac{c}{n'})(v_g-\frac{c}{n})}{\frac{n'}{n}(v_m-v_g)(v_m-\frac{c}{n})+(v_m-\frac{c}{n'})(v_g-\frac{c}{n})} \label{vgformula}
\end{equation}

\begin{figure}
\includegraphics[width=0.68\textwidth]{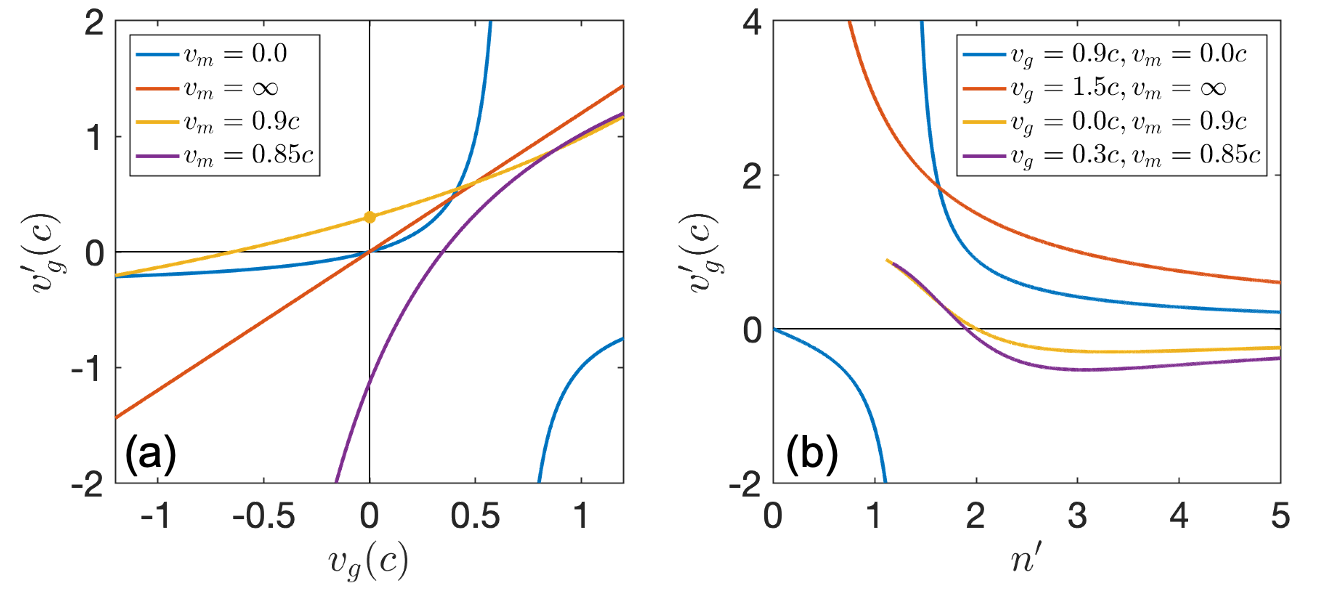}%
\caption{\label{fig_model} Dependency of the final group velocity of the light bullet $v_g'$ on various parameters. (a) Relation of $v_g'$ with initial group velocity $v_g$. For the lines with $v_m$ being $0, \infty, 0.9c$, and $0.85c$, $(n, n')$ are $(2, 1)$, $(2, 1.67)$, $(2, 1.67)$, and $(1.67, 2)$, respectively. (b) Relation of $v_g'$ with the refractive index $n'$ after the boundary. The refractive index before the boundary is $n=2, 2, 2,$ and $1.67$ for the blue, red, yellow, and purple lines, respectively. For the spatiotemporal boundaries represented by the yellow and purple lines, the values of $n'$ are restricted to $n'>c/v_m$.}
\end{figure}

We illustrate the properties of Eq.~(\ref{vgformula}) in Fig.~\ref{fig_model}. Figure \ref{fig_model}a plots $v_g’$ as a function of $v_g$ for various choices of $v_m$, $n$ and $n’$. When $v_m\rightarrow0$, Eq.~(\ref{vgformula}) reduces to $v_g'=\frac{n'cv_g}{(n'^2-n^2)v_g+nc}$. It describes the interaction of a light bullet with a spatial boundary \cite{bhaduri2020anomalous} and is represented by the blue line in Fig.~\ref{fig_model}a. When $v_m\rightarrow\infty$, Eq.~(\ref{vgformula}) reduces to $v_g'=\frac{n}{n'}v_g$. It describes the interaction of a light bullet with a temporal boundary and is represented by the red line in Fig.~\ref{fig_model}a. In both cases, $v_g'=0$ if and only if $v_g=0$. Since $v_g=0$ describes a monochromatic beam, this implies that a monochromatic beam remains monochromatic after interaction with spatial or temporal boundaries.

In the case of spatiotemporal boundaries where $v_m \ne 0$ or $\infty$, it is possible to have $v_g'\neq 0$ while $v_g=0$, or $v_g'=0$ while $v_g\neq 0$, as shown in the yellow and purple lines in Fig.~\ref{fig_model}a where $(v_m, n, n')$ are $(0.9c, 2, 1.67)$ and $(0.85c, 1.67, 2)$ respectively. Therefore, by using spatiotemporal refraction, one can generate a light bullet that moves at a nonzero velocity from a monochromatic focused beam. We call this phenomenon the launching of light bullets. One can also convert a moving light bullet to a monochromatic focused beam that has a stationary intensity distribution. We call this phenomenon the immobilization of light bullets.

In Fig.~\ref{fig_model}b, we plot the dependency of $v_g'$ on $n'$ when $n$, $v_m$, and $v_g$ are fixed. For the temporal boundary, $v_g'$ is always positive (red line in Fig.~\ref{fig_model}b). For the spatial boundary, its interaction with light bullets is known to display anomalous refraction, where the group velocity of a light bullet may increase when entering a high index medium from a low index medium \cite{bhaduri2020anomalous}. Despite displaying the anomalous refraction, $v_g'$ does not go to zero for all positive values of $n'$. However, it is possible to have $v_g'=0$ at some positive value of $n'$ for spatiotemporal boundaries (yellow and purple lines in Fig.~\ref{fig_model}b). When we choose $n'$ to be smaller or larger than this value, $v_g'$ can be either positive or negative, respectively. Thus, from a monochromatic beam, one can generate either a forward-moving or a backward-moving light bullet by properly choosing $n'$ (the yellow line in Fig.~\ref{fig_model}b). For a light bullet that is initially moving, one can change its velocity, immobilize it, or reverse the sign of its velocity also by properly choosing $n'$ (the purple line in Fig.~\ref{fig_model}b).

In the theoretical treatment above we have focused only on the properties of initial and final states of the wavepackets. (We note the initial and final wavepackets, by design, have different characteristics.) We now proceed to numerically demonstrate the entire dynamics of light bullets interacting with a spatiotemporal boundary. Throughout this Letter, we only consider two-dimensional, TM polarized fields (with nonzero field components being the electric field $E_y$ and the magnetic field components $H_x$ and $H_z$). Our methods and findings can be straightforwardly applied to different polarizations and to three-dimensional systems. We first demonstrate the launching of a light bullet from a monochromatic beam. We consider a monochromatic beam in a homogeneous medium of refractive index $n=2$ that has a focal spot at $z=0$. We apply an index front moving in the $+z$-direction at a velocity of $v_m=0.9c$ that changes the index $n$ from $2$ to $1.67$ adiabatically. The index distribution is given by:
\begin{equation}\label{index1}
n'(q) = \begin{dcases}
    2 &(q<70\lambda_0) \\
    \frac{1}{0.55 + 0.05\sin\left[\frac{\pi(q-150\lambda_0)}{160\lambda_0}\right]} &(70\lambda_0 \leq q < 230\lambda_0)\\
    \frac{5}{3} &(230\lambda_0 \leq q)
\end{dcases}
\end{equation}
Here $q = -z\sin\phi+ct\cos\phi$ is the coordinate relative to the moving front. $\phi = \arctan (c/v_m)$. $\lambda_0$ is a reference wavelength, given by the wavelength of the normally incident plane wave component ($k_x=0$) in the medium before the boundary. Such a definition of $\lambda_0$ is used consistently throughout the Letter.

In our numerical simulations, we decompose a light bullet into a superposition of plane waves that satisfy Eqs.~(\ref{lightcone}) and (\ref{stcorrelation}). The behavior of each plane wave under a spatiotemporal boundary is then solved, and their superposition gives rise to the dynamics of the light bullet under the spatiotemporal boundary. The details of the simulation are discussed in the supplementary material \cite{suppmat}.

\begin{figure}
\includegraphics[width=0.68\textwidth]{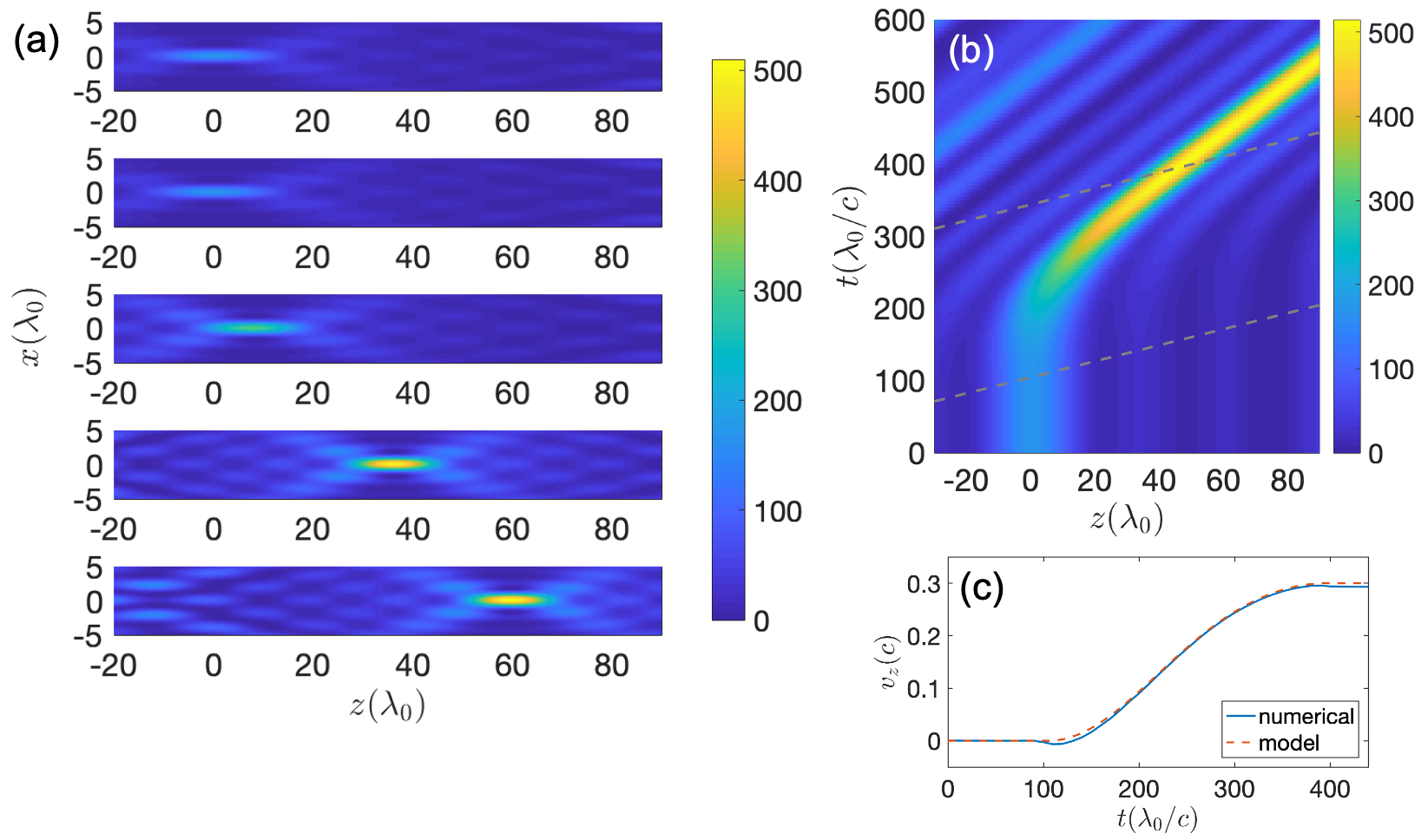}%
\caption{\label{fig_generation} Generation of a light bullet from a monochromatic beam. (a) Distribution of $|E_y|^2$ on the $xz$-plane at time $t=(0, 80, 240, 360, 440)\lambda_0/c$, from top to bottom, respectively. In simulation, we choose $|k_x|<0.26k_0$, where $k_0=2\pi/\lambda_0$. (b) Distribution of $|E_y|^2$ on the $zt$-plane. Gray dashed lines represent the boundaries between which the index varies. Lines are given by $q=70\lambda_0$ and $q=230\lambda_0$. (c) Changes in the group velocity of the light bullet over time.}
\end{figure}

The simulated intensity distribution $|E_y|^2$ on the $xz$-plane at different times $t$ is shown in Fig.~\ref{fig_generation}a. Initially, the focal point of a monochromatic beam is stationary in space. The spatiotemporal boundary causes the point of maximum intensity to move in the $z$-direction and reaches a final velocity of $v_g'=0.294c$. The point $(v_g, v_g')$ is illustrated as the yellow dot in Fig.~\ref{fig_model}a. Such a process can also be visualized in the intensity distribution on the $zt$-plane, shown in Fig.~\ref{fig_generation}b. The two gray dashed lines denote the spatiotemporal boundaries between which the index varies. The regions below the lower line and above the top line have homogeneous indices of $2$ and $1.67$, respectively. The field distribution in $zt$-plane shows a spatiotemporal refraction across the boundary. By numerically determining the point of maximal intensity at different times, one can obtain the velocity of the light bullet as a function of time, shown as the blue line in Fig.~\ref{fig_generation}c. We see that the velocity changes from zero to a final positive value. This demonstrates the launching of a light bullet from a monochromatic focused beam.

The space-time trajectory of the light bullet can be understood using a simple model. Since the process is adiabatic, the instantaneous group velocity, which we also denote as $v_g'$, can be described by Eq.~(\ref{vgformula}) if we interpret $n'$ as the local refractive index at the location of the peak intensity. This allows one to write down a simple model for the space-time trajectory of the light bullet:
\begin{equation}
    \frac{\mathrm{d}z}{\mathrm{d}t}=v_g'(v_g, n, n'(z, t), v_m) \label{zt_traj}
\end{equation}
Here, $n'(z, t)$ is the refractive index distribution of a spatiotemporal boundary (Eq.~(\ref{index1})). All other parameters take fixed values. We plot $v_g'(t)$ calculated from Eq.~(\ref{zt_traj}), shown by the red dashed line in Fig.~\ref{fig_generation}c. Its agreement with our previous numerical simulation is excellent. The reasons for the slight deviation between the two are discussed in the supplementary material \cite{suppmat}.

Now we proceed to demonstrate other control achievable by spatiotemporal boundaries, namely, the programmable motion of light bullets and the immobilization of light bullets. We consider a light bullet in a homogeneous medium of index $n=1.67$ with initial velocity $v_g=0.3c$. We assume the index front moves at a velocity $v_m=0.85c$. From Eq.~(\ref{vgformula}), we see that the instantaneous and final group velocity of the light bullet can be designed by choosing the index profile of the boundary and the index after the boundary, respectively. Specifically, we note that if we increase the index to $n'>1.9$, $v_g'$ becomes negative, and if $n'=1.9$, $v_g'=0$. This dependency is shown as the purple line in Fig.~\ref{fig_model}b.

\begin{figure}
\includegraphics[width=0.68\textwidth]{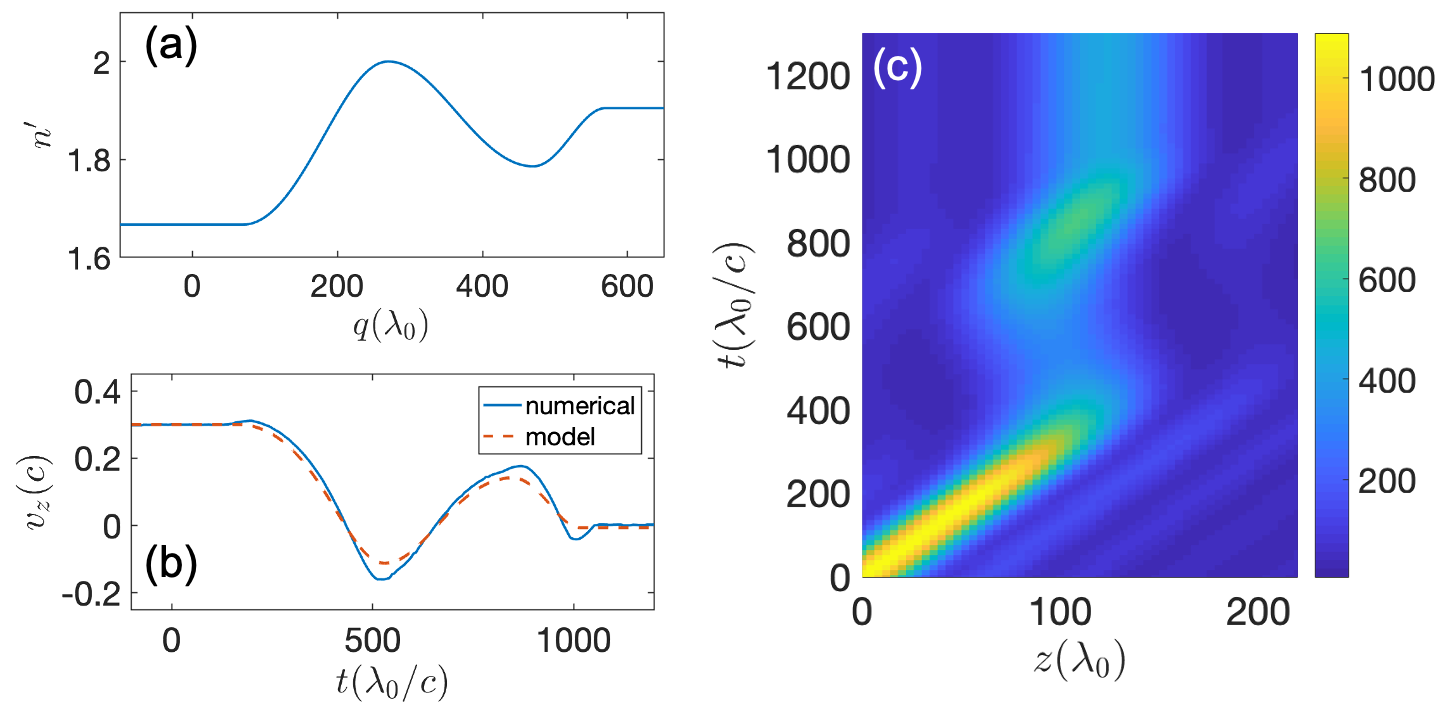}%
\caption{\label{fig_programmable} Programmable motion of a light bullet. (a) The index profile of the spatiotemporal boundary. $q$ is similarly defined as the previous example, except using $v_m=0.85c$. (b) The group velocity of the light bullet as a function of time. (c) The intensity distribution $|E_y|^2$ on the $zt$-plane. The light bullet consists of plane waves with $|k_x|<0.14k_0$.}
\end{figure}

As a numerical demonstration, we consider the index profile of the spatiotemporal boundary shown in Fig.~\ref{fig_programmable}a. A light bullet with $v_g=0.3c$ initially locates at $z=0$ when $t=0$. When the light bullet traverses the spatiotemporal boundary, the numerically simulated intensity distribution on the $zt$-plane is shown in Fig.~\ref{fig_programmable}c. We see the trajectory of the light bullet display a back-and-forth motion. We numerically calculate the group velocity of the light bullet, shown as the blue line in Fig.~\ref{fig_programmable}b. We see that this line has the same shape as the refractive index profile (Fig.~\ref{fig_programmable}a). This shows our ability to directly control the group velocity of the light bullet by designing the index profile of a spatiotemporal boundary, and therefore generate programmable motion of the light bullet. The index after the boundary is chosen to be $n'=1.9$. We see the intensity distribution becomes stationary which demonstrates the immobilization of the light bullet. The behavior of the group velocity can also be described by the model of Eq.~(\ref{zt_traj}). The result is shown as the red dashed line in Fig.~\ref{fig_programmable}b. It reaches a good agreement with the numerically simulated group velocity.

Before concluding the Letter, there are a few technical details worth discussing. Monochromatic focused beams and light bullets, and light bullets with different group velocities are both mathematically connected by Lorentz transformations \cite{yessenov2023relativistic}. Here we propose a physical process that converts monochromatic beams to light bullets, light bullets to monochromatic beams, or light bullets with one group velocity to one with another group velocity.

Experimental demonstration of our predicted effects requires the creation of a refractive index front with substantial refractive index contrast. It should also move at a speed comparable to that of light in vacuum. These requirements can be satisfied in the microwave frequency range \cite{wilson2011observation, moussa2023observation}. In the optical frequency range, it is possible to have an index front that moves at both subluminal and superluminal speed, as has been experimentally implemented \cite{hashimshony2001conversion, philbin2008fiber, fan2023ultrafast}. However, having a substantial index contrast on two sides of the front is more challenging. Nevertheless, we believe the effects we predict may become experimentally feasible in the optical frequency range with the rapid advance of experimental demonstration of time-varying media \cite{hashimshony2001conversion, preble2007changing, tanabe2009dynamic, kampfrath2010ultrafast, philbin2008fiber, vezzoli2018optical, zhou2020broadband, fan2023ultrafast, bacot2016time}.

In this Letter, we have assumed a linear, dispersionless, and lossless medium. This assumption requires us to limit our theory to the case of $v_m>c/n^*$ or $v_m<c/n^*$ (yellow and purple lines in Fig.~\ref{fig_model}b), with $n^*$ being any refractive index along the profile of the spatiotemporal boundary. In the opposite case where at some point $v_m=c/n^*$ is satisfied, we encounter a point known as the optical analog of the event horizon \cite{philbin2008fiber, biancalana2007dynamics}. As we discuss in the supplementary material, both the amplitude and frequency of the electromagnetic wave diverge at this point \cite{suppmat}. Therefore, our assumption of a linear, dispersionless, and lossless medium no longer holds. Novel phenomena induced by such a singularity might be of interest for future studies \cite{de1992optical, philbin2008fiber}. We note that the predicted light bullet behavior can also occur in lossy and dispersive media \cite{suppmat}.

In conclusion, we have demonstrated the unusual capabilities of a spatiotemporal boundary in controlling the space-time trajectory of a light bullet. In particular, we proposed a physical mechanism to launch light bullets from monochromatic focused beams. This may offer a practical way of generating light bullets. By designing the index profile across the spatiotemporal boundary, we can program the group velocity evolution of a light bullet and, therefore, control its trajectory in space-time. We numerically demonstrated such programmable motion and the immobilization of the light bullet. Such physical mechanisms may be utilized to engineer the dynamics of electromagnetic waves in spacetime and therefore may find use in novel spatiotemporal control of material properties or the dynamics of moving particles. Given the ability to arbitrarily engineer the group velocity of light bullets, spatiotemporal refraction of light bullets may also be used as a platform for the laboratory emulation of relativistic physics.

\begin{acknowledgments}
The authors thank Dongha Kim for helpful discussions. This work is supported by a grant from the U. S. Army Research Office (Grant No. W911NF-24-2-0170).
\end{acknowledgments}

\bibliography{ref}

\begin{thebibliography}{64}%
\makeatletter
\providecommand \@ifxundefined [1]{%
 \@ifx{#1\undefined}
}%
\providecommand \@ifnum [1]{%
 \ifnum #1\expandafter \@firstoftwo
 \else \expandafter \@secondoftwo
 \fi
}%
\providecommand \@ifx [1]{%
 \ifx #1\expandafter \@firstoftwo
 \else \expandafter \@secondoftwo
 \fi
}%
\providecommand \natexlab [1]{#1}%
\providecommand \enquote  [1]{``#1''}%
\providecommand \bibnamefont  [1]{#1}%
\providecommand \bibfnamefont [1]{#1}%
\providecommand \citenamefont [1]{#1}%
\providecommand \href@noop [0]{\@secondoftwo}%
\providecommand \href [0]{\begingroup \@sanitize@url \@href}%
\providecommand \@href[1]{\@@startlink{#1}\@@href}%
\providecommand \@@href[1]{\endgroup#1\@@endlink}%
\providecommand \@sanitize@url [0]{\catcode `\\12\catcode `\$12\catcode `\&12\catcode `\#12\catcode `\^12\catcode `\_12\catcode `\%12\relax}%
\providecommand \@@startlink[1]{}%
\providecommand \@@endlink[0]{}%
\providecommand \url  [0]{\begingroup\@sanitize@url \@url }%
\providecommand \@url [1]{\endgroup\@href {#1}{\urlprefix }}%
\providecommand \urlprefix  [0]{URL }%
\providecommand \Eprint [0]{\href }%
\providecommand \doibase [0]{https://doi.org/}%
\providecommand \selectlanguage [0]{\@gobble}%
\providecommand \bibinfo  [0]{\@secondoftwo}%
\providecommand \bibfield  [0]{\@secondoftwo}%
\providecommand \translation [1]{[#1]}%
\providecommand \BibitemOpen [0]{}%
\providecommand \bibitemStop [0]{}%
\providecommand \bibitemNoStop [0]{.\EOS\space}%
\providecommand \EOS [0]{\spacefactor3000\relax}%
\providecommand \BibitemShut  [1]{\csname bibitem#1\endcsname}%
\let\auto@bib@innerbib\@empty
%</preamble>
\bibitem [{\citenamefont {Pendry}(2000)}]{pendry2000negative}%
  \BibitemOpen
  \bibfield  {author} {\bibinfo {author} {\bibfnamefont {J.~B.}\ \bibnamefont {Pendry}},\ }\bibfield  {title} {\bibinfo {title} {Negative refraction makes a perfect lens},\ }\href@noop {} {\bibfield  {journal} {\bibinfo  {journal} {Phys. Rev. Lett.}\ }\textbf {\bibinfo {volume} {85}},\ \bibinfo {pages} {3966} (\bibinfo {year} {2000})}\BibitemShut {NoStop}%
\bibitem [{\citenamefont {Efremidis}\ \emph {et~al.}(2019)\citenamefont {Efremidis}, \citenamefont {Chen}, \citenamefont {Segev},\ and\ \citenamefont {Christodoulides}}]{efremidis2019airy}%
  \BibitemOpen
  \bibfield  {author} {\bibinfo {author} {\bibfnamefont {N.~K.}\ \bibnamefont {Efremidis}}, \bibinfo {author} {\bibfnamefont {Z.}~\bibnamefont {Chen}}, \bibinfo {author} {\bibfnamefont {M.}~\bibnamefont {Segev}},\ and\ \bibinfo {author} {\bibfnamefont {D.~N.}\ \bibnamefont {Christodoulides}},\ }\bibfield  {title} {\bibinfo {title} {Airy beams and accelerating waves: an overview of recent advances},\ }\href@noop {} {\bibfield  {journal} {\bibinfo  {journal} {Optica}\ }\textbf {\bibinfo {volume} {6}},\ \bibinfo {pages} {686} (\bibinfo {year} {2019})}\BibitemShut {NoStop}%
\bibitem [{\citenamefont {Dorrah}\ \emph {et~al.}(2023)\citenamefont {Dorrah}, \citenamefont {Bordoloi}, \citenamefont {de~Angelis}, \citenamefont {de~Sarro}, \citenamefont {Ambrosio}, \citenamefont {Zamboni-Rached},\ and\ \citenamefont {Capasso}}]{dorrah2023light}%
  \BibitemOpen
  \bibfield  {author} {\bibinfo {author} {\bibfnamefont {A.~H.}\ \bibnamefont {Dorrah}}, \bibinfo {author} {\bibfnamefont {P.}~\bibnamefont {Bordoloi}}, \bibinfo {author} {\bibfnamefont {V.~S.}\ \bibnamefont {de~Angelis}}, \bibinfo {author} {\bibfnamefont {J.~O.}\ \bibnamefont {de~Sarro}}, \bibinfo {author} {\bibfnamefont {L.~A.}\ \bibnamefont {Ambrosio}}, \bibinfo {author} {\bibfnamefont {M.}~\bibnamefont {Zamboni-Rached}},\ and\ \bibinfo {author} {\bibfnamefont {F.}~\bibnamefont {Capasso}},\ }\bibfield  {title} {\bibinfo {title} {Light sheets for continuous-depth holography and three-dimensional volumetric displays},\ }\href@noop {} {\bibfield  {journal} {\bibinfo  {journal} {Nature Photonics}\ }\textbf {\bibinfo {volume} {17}},\ \bibinfo {pages} {427} (\bibinfo {year} {2023})}\BibitemShut {NoStop}%
\bibitem [{\citenamefont {Joannopoulos}\ \emph {et~al.}(2008)\citenamefont {Joannopoulos}, \citenamefont {Johnson}, \citenamefont {Winn},\ and\ \citenamefont {Meade}}]{JoannopoulosJohnsonWinnMeade+2008}%
  \BibitemOpen
  \bibfield  {author} {\bibinfo {author} {\bibfnamefont {J.~D.}\ \bibnamefont {Joannopoulos}}, \bibinfo {author} {\bibfnamefont {S.~G.}\ \bibnamefont {Johnson}}, \bibinfo {author} {\bibfnamefont {J.~N.}\ \bibnamefont {Winn}},\ and\ \bibinfo {author} {\bibfnamefont {R.~D.}\ \bibnamefont {Meade}},\ }\href {https://doi.org/doi:10.1515/9781400828241} {\emph {\bibinfo {title} {Photonic Crystals}}}\ (\bibinfo  {publisher} {Princeton University Press},\ \bibinfo {address} {Princeton},\ \bibinfo {year} {2008})\BibitemShut {NoStop}%
\bibitem [{\citenamefont {Yu}\ and\ \citenamefont {Capasso}(2014)}]{yu2014flat}%
  \BibitemOpen
  \bibfield  {author} {\bibinfo {author} {\bibfnamefont {N.}~\bibnamefont {Yu}}\ and\ \bibinfo {author} {\bibfnamefont {F.}~\bibnamefont {Capasso}},\ }\bibfield  {title} {\bibinfo {title} {Flat optics with designer metasurfaces},\ }\href@noop {} {\bibfield  {journal} {\bibinfo  {journal} {Nature materials}\ }\textbf {\bibinfo {volume} {13}},\ \bibinfo {pages} {139} (\bibinfo {year} {2014})}\BibitemShut {NoStop}%
\bibitem [{\citenamefont {Shaltout}\ \emph {et~al.}(2019)\citenamefont {Shaltout}, \citenamefont {Shalaev},\ and\ \citenamefont {Brongersma}}]{shaltout2019spatiotemporal}%
  \BibitemOpen
  \bibfield  {author} {\bibinfo {author} {\bibfnamefont {A.~M.}\ \bibnamefont {Shaltout}}, \bibinfo {author} {\bibfnamefont {V.~M.}\ \bibnamefont {Shalaev}},\ and\ \bibinfo {author} {\bibfnamefont {M.~L.}\ \bibnamefont {Brongersma}},\ }\bibfield  {title} {\bibinfo {title} {Spatiotemporal light control with active metasurfaces},\ }\href@noop {} {\bibfield  {journal} {\bibinfo  {journal} {Science}\ }\textbf {\bibinfo {volume} {364}},\ \bibinfo {pages} {eaat3100} (\bibinfo {year} {2019})}\BibitemShut {NoStop}%
\bibitem [{\citenamefont {Wang}\ \emph {et~al.}(2020)\citenamefont {Wang}, \citenamefont {Guo}, \citenamefont {Zhao},\ and\ \citenamefont {Fan}}]{wang2020compact}%
  \BibitemOpen
  \bibfield  {author} {\bibinfo {author} {\bibfnamefont {H.}~\bibnamefont {Wang}}, \bibinfo {author} {\bibfnamefont {C.}~\bibnamefont {Guo}}, \bibinfo {author} {\bibfnamefont {Z.}~\bibnamefont {Zhao}},\ and\ \bibinfo {author} {\bibfnamefont {S.}~\bibnamefont {Fan}},\ }\bibfield  {title} {\bibinfo {title} {Compact incoherent image differentiation with nanophotonic structures},\ }\href@noop {} {\bibfield  {journal} {\bibinfo  {journal} {ACS Photonics}\ }\textbf {\bibinfo {volume} {7}},\ \bibinfo {pages} {338} (\bibinfo {year} {2020})}\BibitemShut {NoStop}%
\bibitem [{\citenamefont {Wang}\ \emph {et~al.}(2021)\citenamefont {Wang}, \citenamefont {Guo}, \citenamefont {Jin}, \citenamefont {Song},\ and\ \citenamefont {Fan}}]{wang2021engineering}%
  \BibitemOpen
  \bibfield  {author} {\bibinfo {author} {\bibfnamefont {H.}~\bibnamefont {Wang}}, \bibinfo {author} {\bibfnamefont {C.}~\bibnamefont {Guo}}, \bibinfo {author} {\bibfnamefont {W.}~\bibnamefont {Jin}}, \bibinfo {author} {\bibfnamefont {A.~Y.}\ \bibnamefont {Song}},\ and\ \bibinfo {author} {\bibfnamefont {S.}~\bibnamefont {Fan}},\ }\bibfield  {title} {\bibinfo {title} {Engineering arbitrarily oriented spatiotemporal optical vortices using transmission nodal lines},\ }\href@noop {} {\bibfield  {journal} {\bibinfo  {journal} {Optica}\ }\textbf {\bibinfo {volume} {8}},\ \bibinfo {pages} {966} (\bibinfo {year} {2021})}\BibitemShut {NoStop}%
\bibitem [{\citenamefont {Wang}\ \emph {et~al.}(2022)\citenamefont {Wang}, \citenamefont {Jin}, \citenamefont {Guo}, \citenamefont {Zhao}, \citenamefont {Rodrigues},\ and\ \citenamefont {Fan}}]{wang2022design}%
  \BibitemOpen
  \bibfield  {author} {\bibinfo {author} {\bibfnamefont {H.}~\bibnamefont {Wang}}, \bibinfo {author} {\bibfnamefont {W.}~\bibnamefont {Jin}}, \bibinfo {author} {\bibfnamefont {C.}~\bibnamefont {Guo}}, \bibinfo {author} {\bibfnamefont {N.}~\bibnamefont {Zhao}}, \bibinfo {author} {\bibfnamefont {S.~P.}\ \bibnamefont {Rodrigues}},\ and\ \bibinfo {author} {\bibfnamefont {S.}~\bibnamefont {Fan}},\ }\bibfield  {title} {\bibinfo {title} {Design of compact meta-crystal slab for general optical convolution},\ }\href@noop {} {\bibfield  {journal} {\bibinfo  {journal} {ACS Photonics}\ }\textbf {\bibinfo {volume} {9}},\ \bibinfo {pages} {1358} (\bibinfo {year} {2022})}\BibitemShut {NoStop}%
\bibitem [{\citenamefont {Morgenthaler}(1958)}]{morgenthaler1958velocity}%
  \BibitemOpen
  \bibfield  {author} {\bibinfo {author} {\bibfnamefont {F.~R.}\ \bibnamefont {Morgenthaler}},\ }\bibfield  {title} {\bibinfo {title} {Velocity modulation of electromagnetic waves},\ }\href@noop {} {\bibfield  {journal} {\bibinfo  {journal} {IRE Transactions on microwave theory and techniques}\ }\textbf {\bibinfo {volume} {6}},\ \bibinfo {pages} {167} (\bibinfo {year} {1958})}\BibitemShut {NoStop}%
\bibitem [{\citenamefont {Mendon{\c{c}}a}\ \emph {et~al.}(2003)\citenamefont {Mendon{\c{c}}a}, \citenamefont {Martins},\ and\ \citenamefont {Guerreiro}}]{mendoncca2003temporal}%
  \BibitemOpen
  \bibfield  {author} {\bibinfo {author} {\bibfnamefont {J.~T.}\ \bibnamefont {Mendon{\c{c}}a}}, \bibinfo {author} {\bibfnamefont {A.~M.}\ \bibnamefont {Martins}},\ and\ \bibinfo {author} {\bibfnamefont {A.}~\bibnamefont {Guerreiro}},\ }\bibfield  {title} {\bibinfo {title} {Temporal beam splitter and temporal interference},\ }\href@noop {} {\bibfield  {journal} {\bibinfo  {journal} {Phys. Rev. A}\ }\textbf {\bibinfo {volume} {68}},\ \bibinfo {pages} {043801} (\bibinfo {year} {2003})}\BibitemShut {NoStop}%
\bibitem [{\citenamefont {Yanik}\ and\ \citenamefont {Fan}(2004{\natexlab{a}})}]{yanik2004stopping}%
  \BibitemOpen
  \bibfield  {author} {\bibinfo {author} {\bibfnamefont {M.~F.}\ \bibnamefont {Yanik}}\ and\ \bibinfo {author} {\bibfnamefont {S.}~\bibnamefont {Fan}},\ }\bibfield  {title} {\bibinfo {title} {Stopping light all optically},\ }\href@noop {} {\bibfield  {journal} {\bibinfo  {journal} {Phys. Rev. Lett.}\ }\textbf {\bibinfo {volume} {92}},\ \bibinfo {pages} {083901} (\bibinfo {year} {2004}{\natexlab{a}})}\BibitemShut {NoStop}%
\bibitem [{\citenamefont {Yanik}\ and\ \citenamefont {Fan}(2004{\natexlab{b}})}]{yanik2004time}%
  \BibitemOpen
  \bibfield  {author} {\bibinfo {author} {\bibfnamefont {M.~F.}\ \bibnamefont {Yanik}}\ and\ \bibinfo {author} {\bibfnamefont {S.}~\bibnamefont {Fan}},\ }\bibfield  {title} {\bibinfo {title} {Time reversal of light with linear optics and modulators},\ }\href@noop {} {\bibfield  {journal} {\bibinfo  {journal} {Phys. Rev. Lett.}\ }\textbf {\bibinfo {volume} {93}},\ \bibinfo {pages} {173903} (\bibinfo {year} {2004}{\natexlab{b}})}\BibitemShut {NoStop}%
\bibitem [{\citenamefont {Xiao}\ \emph {et~al.}(2014)\citenamefont {Xiao}, \citenamefont {Maywar},\ and\ \citenamefont {Agrawal}}]{xiao2014reflection}%
  \BibitemOpen
  \bibfield  {author} {\bibinfo {author} {\bibfnamefont {Y.}~\bibnamefont {Xiao}}, \bibinfo {author} {\bibfnamefont {D.~N.}\ \bibnamefont {Maywar}},\ and\ \bibinfo {author} {\bibfnamefont {G.~P.}\ \bibnamefont {Agrawal}},\ }\bibfield  {title} {\bibinfo {title} {Reflection and transmission of electromagnetic waves at a temporal boundary},\ }\href@noop {} {\bibfield  {journal} {\bibinfo  {journal} {Opt. Lett.}\ }\textbf {\bibinfo {volume} {39}},\ \bibinfo {pages} {574} (\bibinfo {year} {2014})}\BibitemShut {NoStop}%
\bibitem [{\citenamefont {Plansinis}\ \emph {et~al.}(2015)\citenamefont {Plansinis}, \citenamefont {Donaldson},\ and\ \citenamefont {Agrawal}}]{plansinis2015temporal}%
  \BibitemOpen
  \bibfield  {author} {\bibinfo {author} {\bibfnamefont {B.~W.}\ \bibnamefont {Plansinis}}, \bibinfo {author} {\bibfnamefont {W.~R.}\ \bibnamefont {Donaldson}},\ and\ \bibinfo {author} {\bibfnamefont {G.~P.}\ \bibnamefont {Agrawal}},\ }\bibfield  {title} {\bibinfo {title} {What is the temporal analog of reflection and refraction of optical beams?},\ }\href@noop {} {\bibfield  {journal} {\bibinfo  {journal} {Phys. Rev. Lett.}\ }\textbf {\bibinfo {volume} {115}},\ \bibinfo {pages} {183901} (\bibinfo {year} {2015})}\BibitemShut {NoStop}%
\bibitem [{\citenamefont {Yuan}\ \emph {et~al.}(2018)\citenamefont {Yuan}, \citenamefont {Lin}, \citenamefont {Xiao},\ and\ \citenamefont {Fan}}]{yuan2018synthetic}%
  \BibitemOpen
  \bibfield  {author} {\bibinfo {author} {\bibfnamefont {L.}~\bibnamefont {Yuan}}, \bibinfo {author} {\bibfnamefont {Q.}~\bibnamefont {Lin}}, \bibinfo {author} {\bibfnamefont {M.}~\bibnamefont {Xiao}},\ and\ \bibinfo {author} {\bibfnamefont {S.}~\bibnamefont {Fan}},\ }\bibfield  {title} {\bibinfo {title} {Synthetic dimension in photonics},\ }\href@noop {} {\bibfield  {journal} {\bibinfo  {journal} {Optica}\ }\textbf {\bibinfo {volume} {5}},\ \bibinfo {pages} {1396} (\bibinfo {year} {2018})}\BibitemShut {NoStop}%
\bibitem [{\citenamefont {Pacheco-Pe{\~n}a}\ and\ \citenamefont {Engheta}(2020{\natexlab{a}})}]{pacheco2020temporal}%
  \BibitemOpen
  \bibfield  {author} {\bibinfo {author} {\bibfnamefont {V.}~\bibnamefont {Pacheco-Pe{\~n}a}}\ and\ \bibinfo {author} {\bibfnamefont {N.}~\bibnamefont {Engheta}},\ }\bibfield  {title} {\bibinfo {title} {Temporal aiming},\ }\href@noop {} {\bibfield  {journal} {\bibinfo  {journal} {Light: Science \& Applications}\ }\textbf {\bibinfo {volume} {9}},\ \bibinfo {pages} {129} (\bibinfo {year} {2020}{\natexlab{a}})}\BibitemShut {NoStop}%
\bibitem [{\citenamefont {Pacheco-Pe{\~n}a}\ and\ \citenamefont {Engheta}(2020{\natexlab{b}})}]{pacheco2020antireflection}%
  \BibitemOpen
  \bibfield  {author} {\bibinfo {author} {\bibfnamefont {V.}~\bibnamefont {Pacheco-Pe{\~n}a}}\ and\ \bibinfo {author} {\bibfnamefont {N.}~\bibnamefont {Engheta}},\ }\bibfield  {title} {\bibinfo {title} {Antireflection temporal coatings},\ }\href@noop {} {\bibfield  {journal} {\bibinfo  {journal} {Optica}\ }\textbf {\bibinfo {volume} {7}},\ \bibinfo {pages} {323} (\bibinfo {year} {2020}{\natexlab{b}})}\BibitemShut {NoStop}%
\bibitem [{\citenamefont {Dutt}\ \emph {et~al.}(2020)\citenamefont {Dutt}, \citenamefont {Lin}, \citenamefont {Yuan}, \citenamefont {Minkov}, \citenamefont {Xiao},\ and\ \citenamefont {Fan}}]{dutt2020single}%
  \BibitemOpen
  \bibfield  {author} {\bibinfo {author} {\bibfnamefont {A.}~\bibnamefont {Dutt}}, \bibinfo {author} {\bibfnamefont {Q.}~\bibnamefont {Lin}}, \bibinfo {author} {\bibfnamefont {L.}~\bibnamefont {Yuan}}, \bibinfo {author} {\bibfnamefont {M.}~\bibnamefont {Minkov}}, \bibinfo {author} {\bibfnamefont {M.}~\bibnamefont {Xiao}},\ and\ \bibinfo {author} {\bibfnamefont {S.}~\bibnamefont {Fan}},\ }\bibfield  {title} {\bibinfo {title} {A single photonic cavity with two independent physical synthetic dimensions},\ }\href@noop {} {\bibfield  {journal} {\bibinfo  {journal} {Science}\ }\textbf {\bibinfo {volume} {367}},\ \bibinfo {pages} {59} (\bibinfo {year} {2020})}\BibitemShut {NoStop}%
\bibitem [{\citenamefont {Long}\ \emph {et~al.}(2023)\citenamefont {Long}, \citenamefont {Wang}, \citenamefont {Dutt},\ and\ \citenamefont {Fan}}]{long2023time}%
  \BibitemOpen
  \bibfield  {author} {\bibinfo {author} {\bibfnamefont {O.~Y.}\ \bibnamefont {Long}}, \bibinfo {author} {\bibfnamefont {K.}~\bibnamefont {Wang}}, \bibinfo {author} {\bibfnamefont {A.}~\bibnamefont {Dutt}},\ and\ \bibinfo {author} {\bibfnamefont {S.}~\bibnamefont {Fan}},\ }\bibfield  {title} {\bibinfo {title} {Time reflection and refraction in synthetic frequency dimension},\ }\href@noop {} {\bibfield  {journal} {\bibinfo  {journal} {Phys. Rev. Res.}\ }\textbf {\bibinfo {volume} {5}},\ \bibinfo {pages} {L012046} (\bibinfo {year} {2023})}\BibitemShut {NoStop}%
\bibitem [{\citenamefont {Tirole}\ \emph {et~al.}(2023)\citenamefont {Tirole}, \citenamefont {Vezzoli}, \citenamefont {Galiffi}, \citenamefont {Robertson}, \citenamefont {Maurice}, \citenamefont {Tilmann}, \citenamefont {Maier}, \citenamefont {Pendry},\ and\ \citenamefont {Sapienza}}]{tirole2023double}%
  \BibitemOpen
  \bibfield  {author} {\bibinfo {author} {\bibfnamefont {R.}~\bibnamefont {Tirole}}, \bibinfo {author} {\bibfnamefont {S.}~\bibnamefont {Vezzoli}}, \bibinfo {author} {\bibfnamefont {E.}~\bibnamefont {Galiffi}}, \bibinfo {author} {\bibfnamefont {I.}~\bibnamefont {Robertson}}, \bibinfo {author} {\bibfnamefont {D.}~\bibnamefont {Maurice}}, \bibinfo {author} {\bibfnamefont {B.}~\bibnamefont {Tilmann}}, \bibinfo {author} {\bibfnamefont {S.~A.}\ \bibnamefont {Maier}}, \bibinfo {author} {\bibfnamefont {J.~B.}\ \bibnamefont {Pendry}},\ and\ \bibinfo {author} {\bibfnamefont {R.}~\bibnamefont {Sapienza}},\ }\bibfield  {title} {\bibinfo {title} {Double-slit time diffraction at optical frequencies},\ }\href@noop {} {\bibfield  {journal} {\bibinfo  {journal} {Nature Physics}\ }\textbf {\bibinfo {volume} {19}},\ \bibinfo {pages} {999} (\bibinfo {year} {2023})}\BibitemShut {NoStop}%
\bibitem [{\citenamefont {Lustig}\ \emph {et~al.}(2018)\citenamefont {Lustig}, \citenamefont {Sharabi},\ and\ \citenamefont {Segev}}]{lustig2018topological}%
  \BibitemOpen
  \bibfield  {author} {\bibinfo {author} {\bibfnamefont {E.}~\bibnamefont {Lustig}}, \bibinfo {author} {\bibfnamefont {Y.}~\bibnamefont {Sharabi}},\ and\ \bibinfo {author} {\bibfnamefont {M.}~\bibnamefont {Segev}},\ }\bibfield  {title} {\bibinfo {title} {Topological aspects of photonic time crystals},\ }\href@noop {} {\bibfield  {journal} {\bibinfo  {journal} {Optica}\ }\textbf {\bibinfo {volume} {5}},\ \bibinfo {pages} {1390} (\bibinfo {year} {2018})}\BibitemShut {NoStop}%
\bibitem [{\citenamefont {Lyubarov}\ \emph {et~al.}(2022)\citenamefont {Lyubarov}, \citenamefont {Lumer}, \citenamefont {Dikopoltsev}, \citenamefont {Lustig}, \citenamefont {Sharabi},\ and\ \citenamefont {Segev}}]{lyubarov2022amplified}%
  \BibitemOpen
  \bibfield  {author} {\bibinfo {author} {\bibfnamefont {M.}~\bibnamefont {Lyubarov}}, \bibinfo {author} {\bibfnamefont {Y.}~\bibnamefont {Lumer}}, \bibinfo {author} {\bibfnamefont {A.}~\bibnamefont {Dikopoltsev}}, \bibinfo {author} {\bibfnamefont {E.}~\bibnamefont {Lustig}}, \bibinfo {author} {\bibfnamefont {Y.}~\bibnamefont {Sharabi}},\ and\ \bibinfo {author} {\bibfnamefont {M.}~\bibnamefont {Segev}},\ }\bibfield  {title} {\bibinfo {title} {Amplified emission and lasing in photonic time crystals},\ }\href@noop {} {\bibfield  {journal} {\bibinfo  {journal} {Science}\ }\textbf {\bibinfo {volume} {377}},\ \bibinfo {pages} {425} (\bibinfo {year} {2022})}\BibitemShut {NoStop}%
\bibitem [{\citenamefont {Buddhiraju}\ \emph {et~al.}(2020)\citenamefont {Buddhiraju}, \citenamefont {Li},\ and\ \citenamefont {Fan}}]{buddhiraju2020photonic}%
  \BibitemOpen
  \bibfield  {author} {\bibinfo {author} {\bibfnamefont {S.}~\bibnamefont {Buddhiraju}}, \bibinfo {author} {\bibfnamefont {W.}~\bibnamefont {Li}},\ and\ \bibinfo {author} {\bibfnamefont {S.}~\bibnamefont {Fan}},\ }\bibfield  {title} {\bibinfo {title} {Photonic refrigeration from time-modulated thermal emission},\ }\href@noop {} {\bibfield  {journal} {\bibinfo  {journal} {Phys. Rev. Lett.}\ }\textbf {\bibinfo {volume} {124}},\ \bibinfo {pages} {077402} (\bibinfo {year} {2020})}\BibitemShut {NoStop}%
\bibitem [{\citenamefont {Preble}\ \emph {et~al.}(2007)\citenamefont {Preble}, \citenamefont {Xu},\ and\ \citenamefont {Lipson}}]{preble2007changing}%
  \BibitemOpen
  \bibfield  {author} {\bibinfo {author} {\bibfnamefont {S.~F.}\ \bibnamefont {Preble}}, \bibinfo {author} {\bibfnamefont {Q.}~\bibnamefont {Xu}},\ and\ \bibinfo {author} {\bibfnamefont {M.}~\bibnamefont {Lipson}},\ }\bibfield  {title} {\bibinfo {title} {Changing the colour of light in a silicon resonator},\ }\href@noop {} {\bibfield  {journal} {\bibinfo  {journal} {Nature Photonics}\ }\textbf {\bibinfo {volume} {1}},\ \bibinfo {pages} {293} (\bibinfo {year} {2007})}\BibitemShut {NoStop}%
\bibitem [{\citenamefont {Tanabe}\ \emph {et~al.}(2009)\citenamefont {Tanabe}, \citenamefont {Notomi}, \citenamefont {Taniyama},\ and\ \citenamefont {Kuramochi}}]{tanabe2009dynamic}%
  \BibitemOpen
  \bibfield  {author} {\bibinfo {author} {\bibfnamefont {T.}~\bibnamefont {Tanabe}}, \bibinfo {author} {\bibfnamefont {M.}~\bibnamefont {Notomi}}, \bibinfo {author} {\bibfnamefont {H.}~\bibnamefont {Taniyama}},\ and\ \bibinfo {author} {\bibfnamefont {E.}~\bibnamefont {Kuramochi}},\ }\bibfield  {title} {\bibinfo {title} {Dynamic release of trapped light from an ultrahigh-q nanocavity via adiabatic frequency tuning},\ }\href@noop {} {\bibfield  {journal} {\bibinfo  {journal} {Phys. Rev. Lett.}\ }\textbf {\bibinfo {volume} {102}},\ \bibinfo {pages} {043907} (\bibinfo {year} {2009})}\BibitemShut {NoStop}%
\bibitem [{\citenamefont {Kampfrath}\ \emph {et~al.}(2010)\citenamefont {Kampfrath}, \citenamefont {Beggs}, \citenamefont {White}, \citenamefont {Melloni}, \citenamefont {Krauss},\ and\ \citenamefont {Kuipers}}]{kampfrath2010ultrafast}%
  \BibitemOpen
  \bibfield  {author} {\bibinfo {author} {\bibfnamefont {T.}~\bibnamefont {Kampfrath}}, \bibinfo {author} {\bibfnamefont {D.~M.}\ \bibnamefont {Beggs}}, \bibinfo {author} {\bibfnamefont {T.~P.}\ \bibnamefont {White}}, \bibinfo {author} {\bibfnamefont {A.}~\bibnamefont {Melloni}}, \bibinfo {author} {\bibfnamefont {T.~F.}\ \bibnamefont {Krauss}},\ and\ \bibinfo {author} {\bibfnamefont {L.}~\bibnamefont {Kuipers}},\ }\bibfield  {title} {\bibinfo {title} {Ultrafast adiabatic manipulation of slow light in a photonic crystal},\ }\href@noop {} {\bibfield  {journal} {\bibinfo  {journal} {Phys. Rev. A}\ }\textbf {\bibinfo {volume} {81}},\ \bibinfo {pages} {043837} (\bibinfo {year} {2010})}\BibitemShut {NoStop}%
\bibitem [{\citenamefont {Howard}\ \emph {et~al.}(2019)\citenamefont {Howard}, \citenamefont {Turnbull}, \citenamefont {Davies}, \citenamefont {Franke}, \citenamefont {Froula},\ and\ \citenamefont {Palastro}}]{howard2019photon}%
  \BibitemOpen
  \bibfield  {author} {\bibinfo {author} {\bibfnamefont {A.~J.}\ \bibnamefont {Howard}}, \bibinfo {author} {\bibfnamefont {D.}~\bibnamefont {Turnbull}}, \bibinfo {author} {\bibfnamefont {A.~S.}\ \bibnamefont {Davies}}, \bibinfo {author} {\bibfnamefont {P.}~\bibnamefont {Franke}}, \bibinfo {author} {\bibfnamefont {D.~H.}\ \bibnamefont {Froula}},\ and\ \bibinfo {author} {\bibfnamefont {J.~P.}\ \bibnamefont {Palastro}},\ }\bibfield  {title} {\bibinfo {title} {Photon acceleration in a flying focus},\ }\href@noop {} {\bibfield  {journal} {\bibinfo  {journal} {Phys. Rev. Lett.}\ }\textbf {\bibinfo {volume} {123}},\ \bibinfo {pages} {124801} (\bibinfo {year} {2019})}\BibitemShut {NoStop}%
\bibitem [{\citenamefont {Shcherbakov}\ \emph {et~al.}(2019)\citenamefont {Shcherbakov}, \citenamefont {Werner}, \citenamefont {Fan}, \citenamefont {Talisa}, \citenamefont {Chowdhury},\ and\ \citenamefont {Shvets}}]{shcherbakov2019photon}%
  \BibitemOpen
  \bibfield  {author} {\bibinfo {author} {\bibfnamefont {M.~R.}\ \bibnamefont {Shcherbakov}}, \bibinfo {author} {\bibfnamefont {K.}~\bibnamefont {Werner}}, \bibinfo {author} {\bibfnamefont {Z.}~\bibnamefont {Fan}}, \bibinfo {author} {\bibfnamefont {N.}~\bibnamefont {Talisa}}, \bibinfo {author} {\bibfnamefont {E.}~\bibnamefont {Chowdhury}},\ and\ \bibinfo {author} {\bibfnamefont {G.}~\bibnamefont {Shvets}},\ }\bibfield  {title} {\bibinfo {title} {Photon acceleration and tunable broadband harmonics generation in nonlinear time-dependent metasurfaces},\ }\href@noop {} {\bibfield  {journal} {\bibinfo  {journal} {Nature Communications}\ }\textbf {\bibinfo {volume} {10}},\ \bibinfo {pages} {1345} (\bibinfo {year} {2019})}\BibitemShut {NoStop}%
\bibitem [{\citenamefont {Zhou}\ \emph {et~al.}(2020)\citenamefont {Zhou}, \citenamefont {Alam}, \citenamefont {Karimi}, \citenamefont {Upham}, \citenamefont {Reshef}, \citenamefont {Liu}, \citenamefont {Willner},\ and\ \citenamefont {Boyd}}]{zhou2020broadband}%
  \BibitemOpen
  \bibfield  {author} {\bibinfo {author} {\bibfnamefont {Y.}~\bibnamefont {Zhou}}, \bibinfo {author} {\bibfnamefont {M.~Z.}\ \bibnamefont {Alam}}, \bibinfo {author} {\bibfnamefont {M.}~\bibnamefont {Karimi}}, \bibinfo {author} {\bibfnamefont {J.}~\bibnamefont {Upham}}, \bibinfo {author} {\bibfnamefont {O.}~\bibnamefont {Reshef}}, \bibinfo {author} {\bibfnamefont {C.}~\bibnamefont {Liu}}, \bibinfo {author} {\bibfnamefont {A.~E.}\ \bibnamefont {Willner}},\ and\ \bibinfo {author} {\bibfnamefont {R.~W.}\ \bibnamefont {Boyd}},\ }\bibfield  {title} {\bibinfo {title} {Broadband frequency translation through time refraction in an epsilon-near-zero material},\ }\href@noop {} {\bibfield  {journal} {\bibinfo  {journal} {Nature Communications}\ }\textbf {\bibinfo {volume} {11}},\ \bibinfo {pages} {2180} (\bibinfo {year} {2020})}\BibitemShut {NoStop}%
\bibitem [{\citenamefont {Bacot}\ \emph {et~al.}(2016)\citenamefont {Bacot}, \citenamefont {Labousse}, \citenamefont {Eddi}, \citenamefont {Fink},\ and\ \citenamefont {Fort}}]{bacot2016time}%
  \BibitemOpen
  \bibfield  {author} {\bibinfo {author} {\bibfnamefont {V.}~\bibnamefont {Bacot}}, \bibinfo {author} {\bibfnamefont {M.}~\bibnamefont {Labousse}}, \bibinfo {author} {\bibfnamefont {A.}~\bibnamefont {Eddi}}, \bibinfo {author} {\bibfnamefont {M.}~\bibnamefont {Fink}},\ and\ \bibinfo {author} {\bibfnamefont {E.}~\bibnamefont {Fort}},\ }\bibfield  {title} {\bibinfo {title} {Time reversal and holography with spacetime transformations},\ }\href@noop {} {\bibfield  {journal} {\bibinfo  {journal} {Nature Physics}\ }\textbf {\bibinfo {volume} {12}},\ \bibinfo {pages} {972} (\bibinfo {year} {2016})}\BibitemShut {NoStop}%
\bibitem [{\citenamefont {Vezzoli}\ \emph {et~al.}(2018)\citenamefont {Vezzoli}, \citenamefont {Bruno}, \citenamefont {DeVault}, \citenamefont {Roger}, \citenamefont {Shalaev}, \citenamefont {Boltasseva}, \citenamefont {Ferrera}, \citenamefont {Clerici}, \citenamefont {Dubietis},\ and\ \citenamefont {Faccio}}]{vezzoli2018optical}%
  \BibitemOpen
  \bibfield  {author} {\bibinfo {author} {\bibfnamefont {S.}~\bibnamefont {Vezzoli}}, \bibinfo {author} {\bibfnamefont {V.}~\bibnamefont {Bruno}}, \bibinfo {author} {\bibfnamefont {C.}~\bibnamefont {DeVault}}, \bibinfo {author} {\bibfnamefont {T.}~\bibnamefont {Roger}}, \bibinfo {author} {\bibfnamefont {V.~M.}\ \bibnamefont {Shalaev}}, \bibinfo {author} {\bibfnamefont {A.}~\bibnamefont {Boltasseva}}, \bibinfo {author} {\bibfnamefont {M.}~\bibnamefont {Ferrera}}, \bibinfo {author} {\bibfnamefont {M.}~\bibnamefont {Clerici}}, \bibinfo {author} {\bibfnamefont {A.}~\bibnamefont {Dubietis}},\ and\ \bibinfo {author} {\bibfnamefont {D.}~\bibnamefont {Faccio}},\ }\bibfield  {title} {\bibinfo {title} {Optical time reversal from time-dependent epsilon-near-zero media},\ }\href@noop {} {\bibfield  {journal} {\bibinfo  {journal} {Phys. Rev. Lett.}\ }\textbf {\bibinfo {volume} {120}},\ \bibinfo {pages} {043902} (\bibinfo {year} {2018})}\BibitemShut {NoStop}%
\bibitem [{\citenamefont {Moussa}\ \emph {et~al.}(2023)\citenamefont {Moussa}, \citenamefont {Xu}, \citenamefont {Yin}, \citenamefont {Galiffi}, \citenamefont {Ra’di},\ and\ \citenamefont {Al{\`u}}}]{moussa2023observation}%
  \BibitemOpen
  \bibfield  {author} {\bibinfo {author} {\bibfnamefont {H.}~\bibnamefont {Moussa}}, \bibinfo {author} {\bibfnamefont {G.}~\bibnamefont {Xu}}, \bibinfo {author} {\bibfnamefont {S.}~\bibnamefont {Yin}}, \bibinfo {author} {\bibfnamefont {E.}~\bibnamefont {Galiffi}}, \bibinfo {author} {\bibfnamefont {Y.}~\bibnamefont {Ra’di}},\ and\ \bibinfo {author} {\bibfnamefont {A.}~\bibnamefont {Al{\`u}}},\ }\bibfield  {title} {\bibinfo {title} {Observation of temporal reflection and broadband frequency translation at photonic time interfaces},\ }\href@noop {} {\bibfield  {journal} {\bibinfo  {journal} {Nature Physics}\ }\textbf {\bibinfo {volume} {19}},\ \bibinfo {pages} {863} (\bibinfo {year} {2023})}\BibitemShut {NoStop}%
\bibitem [{\citenamefont {Wilson}\ \emph {et~al.}(2011)\citenamefont {Wilson}, \citenamefont {Johansson}, \citenamefont {Pourkabirian}, \citenamefont {Simoen}, \citenamefont {Johansson}, \citenamefont {Duty}, \citenamefont {Nori},\ and\ \citenamefont {Delsing}}]{wilson2011observation}%
  \BibitemOpen
  \bibfield  {author} {\bibinfo {author} {\bibfnamefont {C.~M.}\ \bibnamefont {Wilson}}, \bibinfo {author} {\bibfnamefont {G.}~\bibnamefont {Johansson}}, \bibinfo {author} {\bibfnamefont {A.}~\bibnamefont {Pourkabirian}}, \bibinfo {author} {\bibfnamefont {M.}~\bibnamefont {Simoen}}, \bibinfo {author} {\bibfnamefont {J.~R.}\ \bibnamefont {Johansson}}, \bibinfo {author} {\bibfnamefont {T.}~\bibnamefont {Duty}}, \bibinfo {author} {\bibfnamefont {F.}~\bibnamefont {Nori}},\ and\ \bibinfo {author} {\bibfnamefont {P.}~\bibnamefont {Delsing}},\ }\bibfield  {title} {\bibinfo {title} {Observation of the dynamical casimir effect in a superconducting circuit},\ }\href@noop {} {\bibfield  {journal} {\bibinfo  {journal} {nature}\ }\textbf {\bibinfo {volume} {479}},\ \bibinfo {pages} {376} (\bibinfo {year} {2011})}\BibitemShut {NoStop}%
\bibitem [{\citenamefont {Hashimshony}\ \emph {et~al.}(2001)\citenamefont {Hashimshony}, \citenamefont {Zigler},\ and\ \citenamefont {Papadopoulos}}]{hashimshony2001conversion}%
  \BibitemOpen
  \bibfield  {author} {\bibinfo {author} {\bibfnamefont {D.}~\bibnamefont {Hashimshony}}, \bibinfo {author} {\bibfnamefont {A.}~\bibnamefont {Zigler}},\ and\ \bibinfo {author} {\bibfnamefont {K.}~\bibnamefont {Papadopoulos}},\ }\bibfield  {title} {\bibinfo {title} {Conversion of electrostatic to electromagnetic waves by superluminous ionization fronts},\ }\href@noop {} {\bibfield  {journal} {\bibinfo  {journal} {Phys. Rev. Lett.}\ }\textbf {\bibinfo {volume} {86}},\ \bibinfo {pages} {2806} (\bibinfo {year} {2001})}\BibitemShut {NoStop}%
\bibitem [{\citenamefont {Philbin}\ \emph {et~al.}(2008)\citenamefont {Philbin}, \citenamefont {Kuklewicz}, \citenamefont {Robertson}, \citenamefont {Hill}, \citenamefont {Konig},\ and\ \citenamefont {Leonhardt}}]{philbin2008fiber}%
  \BibitemOpen
  \bibfield  {author} {\bibinfo {author} {\bibfnamefont {T.~G.}\ \bibnamefont {Philbin}}, \bibinfo {author} {\bibfnamefont {C.}~\bibnamefont {Kuklewicz}}, \bibinfo {author} {\bibfnamefont {S.}~\bibnamefont {Robertson}}, \bibinfo {author} {\bibfnamefont {S.}~\bibnamefont {Hill}}, \bibinfo {author} {\bibfnamefont {F.}~\bibnamefont {Konig}},\ and\ \bibinfo {author} {\bibfnamefont {U.}~\bibnamefont {Leonhardt}},\ }\bibfield  {title} {\bibinfo {title} {Fiber-optical analog of the event horizon},\ }\href@noop {} {\bibfield  {journal} {\bibinfo  {journal} {Science}\ }\textbf {\bibinfo {volume} {319}},\ \bibinfo {pages} {1367} (\bibinfo {year} {2008})}\BibitemShut {NoStop}%
\bibitem [{\citenamefont {Biancalana}\ \emph {et~al.}(2007)\citenamefont {Biancalana}, \citenamefont {Amann}, \citenamefont {Uskov},\ and\ \citenamefont {O'Reilly}}]{biancalana2007dynamics}%
  \BibitemOpen
  \bibfield  {author} {\bibinfo {author} {\bibfnamefont {F.}~\bibnamefont {Biancalana}}, \bibinfo {author} {\bibfnamefont {A.}~\bibnamefont {Amann}}, \bibinfo {author} {\bibfnamefont {A.~V.}\ \bibnamefont {Uskov}},\ and\ \bibinfo {author} {\bibfnamefont {E.~P.}\ \bibnamefont {O'Reilly}},\ }\bibfield  {title} {\bibinfo {title} {Dynamics of light propagation in spatiotemporal dielectric structures},\ }\href@noop {} {\bibfield  {journal} {\bibinfo  {journal} {Phys. Rev. E}\ }\textbf {\bibinfo {volume} {75}},\ \bibinfo {pages} {046607} (\bibinfo {year} {2007})}\BibitemShut {NoStop}%
\bibitem [{\citenamefont {Shi}\ and\ \citenamefont {Fan}(2016)}]{shi2016dynamic}%
  \BibitemOpen
  \bibfield  {author} {\bibinfo {author} {\bibfnamefont {Y.}~\bibnamefont {Shi}}\ and\ \bibinfo {author} {\bibfnamefont {S.}~\bibnamefont {Fan}},\ }\bibfield  {title} {\bibinfo {title} {Dynamic non-reciprocal meta-surfaces with arbitrary phase reconfigurability based on photonic transition in meta-atoms},\ }\href@noop {} {\bibfield  {journal} {\bibinfo  {journal} {Applied Physics Letters}\ }\textbf {\bibinfo {volume} {108}} (\bibinfo {year} {2016})}\BibitemShut {NoStop}%
\bibitem [{\citenamefont {Subkhangulov}\ \emph {et~al.}(2016)\citenamefont {Subkhangulov}, \citenamefont {Mikhaylovskiy}, \citenamefont {Zvezdin}, \citenamefont {Kruglyak}, \citenamefont {Rasing},\ and\ \citenamefont {Kimel}}]{subkhangulov2016terahertz}%
  \BibitemOpen
  \bibfield  {author} {\bibinfo {author} {\bibfnamefont {R.}~\bibnamefont {Subkhangulov}}, \bibinfo {author} {\bibfnamefont {R.}~\bibnamefont {Mikhaylovskiy}}, \bibinfo {author} {\bibfnamefont {A.}~\bibnamefont {Zvezdin}}, \bibinfo {author} {\bibfnamefont {V.}~\bibnamefont {Kruglyak}}, \bibinfo {author} {\bibfnamefont {T.}~\bibnamefont {Rasing}},\ and\ \bibinfo {author} {\bibfnamefont {A.}~\bibnamefont {Kimel}},\ }\bibfield  {title} {\bibinfo {title} {Terahertz modulation of the faraday rotation by laser pulses via the optical kerr effect},\ }\href@noop {} {\bibfield  {journal} {\bibinfo  {journal} {Nature Photonics}\ }\textbf {\bibinfo {volume} {10}},\ \bibinfo {pages} {111} (\bibinfo {year} {2016})}\BibitemShut {NoStop}%
\bibitem [{\citenamefont {Guo}\ \emph {et~al.}(2019)\citenamefont {Guo}, \citenamefont {Ding}, \citenamefont {Duan},\ and\ \citenamefont {Ni}}]{guo2019nonreciprocal}%
  \BibitemOpen
  \bibfield  {author} {\bibinfo {author} {\bibfnamefont {X.}~\bibnamefont {Guo}}, \bibinfo {author} {\bibfnamefont {Y.}~\bibnamefont {Ding}}, \bibinfo {author} {\bibfnamefont {Y.}~\bibnamefont {Duan}},\ and\ \bibinfo {author} {\bibfnamefont {X.}~\bibnamefont {Ni}},\ }\bibfield  {title} {\bibinfo {title} {Nonreciprocal metasurface with space--time phase modulation},\ }\href@noop {} {\bibfield  {journal} {\bibinfo  {journal} {Light: Science \& Applications}\ }\textbf {\bibinfo {volume} {8}},\ \bibinfo {pages} {123} (\bibinfo {year} {2019})}\BibitemShut {NoStop}%
\bibitem [{\citenamefont {Sharabi}\ \emph {et~al.}(2022)\citenamefont {Sharabi}, \citenamefont {Dikopoltsev}, \citenamefont {Lustig}, \citenamefont {Lumer},\ and\ \citenamefont {Segev}}]{sharabi2022spatiotemporal}%
  \BibitemOpen
  \bibfield  {author} {\bibinfo {author} {\bibfnamefont {Y.}~\bibnamefont {Sharabi}}, \bibinfo {author} {\bibfnamefont {A.}~\bibnamefont {Dikopoltsev}}, \bibinfo {author} {\bibfnamefont {E.}~\bibnamefont {Lustig}}, \bibinfo {author} {\bibfnamefont {Y.}~\bibnamefont {Lumer}},\ and\ \bibinfo {author} {\bibfnamefont {M.}~\bibnamefont {Segev}},\ }\bibfield  {title} {\bibinfo {title} {Spatiotemporal photonic crystals},\ }\href@noop {} {\bibfield  {journal} {\bibinfo  {journal} {Optica}\ }\textbf {\bibinfo {volume} {9}},\ \bibinfo {pages} {585} (\bibinfo {year} {2022})}\BibitemShut {NoStop}%
\bibitem [{\citenamefont {Fan}\ \emph {et~al.}(2023)\citenamefont {Fan}, \citenamefont {Shaltout}, \citenamefont {van~de Groep}, \citenamefont {Brongersma},\ and\ \citenamefont {Lindenberg}}]{fan2023ultrafast}%
  \BibitemOpen
  \bibfield  {author} {\bibinfo {author} {\bibfnamefont {Q.}~\bibnamefont {Fan}}, \bibinfo {author} {\bibfnamefont {A.~M.}\ \bibnamefont {Shaltout}}, \bibinfo {author} {\bibfnamefont {J.}~\bibnamefont {van~de Groep}}, \bibinfo {author} {\bibfnamefont {M.~L.}\ \bibnamefont {Brongersma}},\ and\ \bibinfo {author} {\bibfnamefont {A.~M.}\ \bibnamefont {Lindenberg}},\ }\bibfield  {title} {\bibinfo {title} {Ultrafast wavefront shaping via space-time refraction},\ }\href@noop {} {\bibfield  {journal} {\bibinfo  {journal} {ACS Photonics}\ }\textbf {\bibinfo {volume} {10}},\ \bibinfo {pages} {2467} (\bibinfo {year} {2023})}\BibitemShut {NoStop}%
\bibitem [{\citenamefont {Caloz}\ and\ \citenamefont {Deck-Leger}(2019)}]{caloz2019spacetime}%
  \BibitemOpen
  \bibfield  {author} {\bibinfo {author} {\bibfnamefont {C.}~\bibnamefont {Caloz}}\ and\ \bibinfo {author} {\bibfnamefont {Z.-L.}\ \bibnamefont {Deck-Leger}},\ }\bibfield  {title} {\bibinfo {title} {Spacetime metamaterials—part ii: Theory and applications},\ }\href@noop {} {\bibfield  {journal} {\bibinfo  {journal} {IEEE Transactions on Antennas and Propagation}\ }\textbf {\bibinfo {volume} {68}},\ \bibinfo {pages} {1583} (\bibinfo {year} {2019})}\BibitemShut {NoStop}%
\bibitem [{\citenamefont {Lu}\ and\ \citenamefont {Greenleaf}(1992)}]{lu1992nondiffracting}%
  \BibitemOpen
  \bibfield  {author} {\bibinfo {author} {\bibfnamefont {J.-Y.}\ \bibnamefont {Lu}}\ and\ \bibinfo {author} {\bibfnamefont {J.~F.}\ \bibnamefont {Greenleaf}},\ }\bibfield  {title} {\bibinfo {title} {Nondiffracting x waves-exact solutions to free-space scalar wave equation and their finite aperture realizations},\ }\href@noop {} {\bibfield  {journal} {\bibinfo  {journal} {IEEE Transactions on Ultrasonics, Ferroelectrics, and Frequency Control}\ }\textbf {\bibinfo {volume} {39}},\ \bibinfo {pages} {19} (\bibinfo {year} {1992})}\BibitemShut {NoStop}%
\bibitem [{\citenamefont {Saari}\ and\ \citenamefont {Reivelt}(1997)}]{saari1997evidence}%
  \BibitemOpen
  \bibfield  {author} {\bibinfo {author} {\bibfnamefont {P.}~\bibnamefont {Saari}}\ and\ \bibinfo {author} {\bibfnamefont {K.}~\bibnamefont {Reivelt}},\ }\bibfield  {title} {\bibinfo {title} {Evidence of x-shaped propagation-invariant localized light waves},\ }\href@noop {} {\bibfield  {journal} {\bibinfo  {journal} {Phys. Rev. Lett.}\ }\textbf {\bibinfo {volume} {79}},\ \bibinfo {pages} {4135} (\bibinfo {year} {1997})}\BibitemShut {NoStop}%
\bibitem [{\citenamefont {Zamboni-Rached}\ \emph {et~al.}(2002)\citenamefont {Zamboni-Rached}, \citenamefont {Recami},\ and\ \citenamefont {Hern{\'a}ndez-Figueroa}}]{zamboni2002new}%
  \BibitemOpen
  \bibfield  {author} {\bibinfo {author} {\bibfnamefont {M.}~\bibnamefont {Zamboni-Rached}}, \bibinfo {author} {\bibfnamefont {E.}~\bibnamefont {Recami}},\ and\ \bibinfo {author} {\bibfnamefont {H.~E.}\ \bibnamefont {Hern{\'a}ndez-Figueroa}},\ }\bibfield  {title} {\bibinfo {title} {New localized superluminal solutions to the wave equations with finite total energies and arbitrary frequencies},\ }\href@noop {} {\bibfield  {journal} {\bibinfo  {journal} {The European Physical Journal D-Atomic, Molecular, Optical and Plasma Physics}\ }\textbf {\bibinfo {volume} {21}},\ \bibinfo {pages} {217} (\bibinfo {year} {2002})}\BibitemShut {NoStop}%
\bibitem [{\citenamefont {Saari}\ and\ \citenamefont {Reivelt}(2004)}]{saari2004generation}%
  \BibitemOpen
  \bibfield  {author} {\bibinfo {author} {\bibfnamefont {P.}~\bibnamefont {Saari}}\ and\ \bibinfo {author} {\bibfnamefont {K.}~\bibnamefont {Reivelt}},\ }\bibfield  {title} {\bibinfo {title} {Generation and classification of localized waves by lorentz transformations in fourier space},\ }\href@noop {} {\bibfield  {journal} {\bibinfo  {journal} {Phys. Rev. E}\ }\textbf {\bibinfo {volume} {69}},\ \bibinfo {pages} {036612} (\bibinfo {year} {2004})}\BibitemShut {NoStop}%
\bibitem [{\citenamefont {Hern{\'a}ndez-Figueroa}\ \emph {et~al.}(2008)\citenamefont {Hern{\'a}ndez-Figueroa}, \citenamefont {Zamboni-Rached},\ and\ \citenamefont {Recami}}]{hernandez2008localized}%
  \BibitemOpen
  \bibfield  {author} {\bibinfo {author} {\bibfnamefont {H.~E.}\ \bibnamefont {Hern{\'a}ndez-Figueroa}}, \bibinfo {author} {\bibfnamefont {M.}~\bibnamefont {Zamboni-Rached}},\ and\ \bibinfo {author} {\bibfnamefont {E.}~\bibnamefont {Recami}},\ }\href@noop {} {\emph {\bibinfo {title} {Localized waves}}},\ Vol.\ \bibinfo {volume} {194}\ (\bibinfo  {publisher} {John Wiley \& Sons},\ \bibinfo {year} {2008})\BibitemShut {NoStop}%
\bibitem [{\citenamefont {Hern{\'a}ndez-Figueroa}\ \emph {et~al.}(2013)\citenamefont {Hern{\'a}ndez-Figueroa}, \citenamefont {Zamboni-Rached},\ and\ \citenamefont {Recami}}]{hernandez2013non}%
  \BibitemOpen
  \bibfield  {author} {\bibinfo {author} {\bibfnamefont {H.~E.}\ \bibnamefont {Hern{\'a}ndez-Figueroa}}, \bibinfo {author} {\bibfnamefont {M.}~\bibnamefont {Zamboni-Rached}},\ and\ \bibinfo {author} {\bibfnamefont {E.}~\bibnamefont {Recami}},\ }\href@noop {} {\emph {\bibinfo {title} {Non-diffracting waves}}}\ (\bibinfo  {publisher} {John Wiley \& Sons},\ \bibinfo {year} {2013})\BibitemShut {NoStop}%
\bibitem [{\citenamefont {Turunen}\ and\ \citenamefont {Friberg}(2010)}]{turunen2010propagation}%
  \BibitemOpen
  \bibfield  {author} {\bibinfo {author} {\bibfnamefont {J.}~\bibnamefont {Turunen}}\ and\ \bibinfo {author} {\bibfnamefont {A.~T.}\ \bibnamefont {Friberg}},\ }\bibfield  {title} {\bibinfo {title} {Propagation-invariant optical fields},\ }in\ \href@noop {} {\emph {\bibinfo {booktitle} {Progress in Optics}}},\ Vol.~\bibinfo {volume} {54}\ (\bibinfo  {publisher} {Elsevier},\ \bibinfo {year} {2010})\ pp.\ \bibinfo {pages} {1--88}\BibitemShut {NoStop}%
\bibitem [{\citenamefont {Kondakci}\ and\ \citenamefont {Abouraddy}(2017)}]{kondakci2017diffraction}%
  \BibitemOpen
  \bibfield  {author} {\bibinfo {author} {\bibfnamefont {H.~E.}\ \bibnamefont {Kondakci}}\ and\ \bibinfo {author} {\bibfnamefont {A.~F.}\ \bibnamefont {Abouraddy}},\ }\bibfield  {title} {\bibinfo {title} {Diffraction-free space--time light sheets},\ }\href@noop {} {\bibfield  {journal} {\bibinfo  {journal} {Nature Photonics}\ }\textbf {\bibinfo {volume} {11}},\ \bibinfo {pages} {733} (\bibinfo {year} {2017})}\BibitemShut {NoStop}%
\bibitem [{\citenamefont {Li}\ and\ \citenamefont {Kawanaka}(2020{\natexlab{a}})}]{li2020velocity}%
  \BibitemOpen
  \bibfield  {author} {\bibinfo {author} {\bibfnamefont {Z.}~\bibnamefont {Li}}\ and\ \bibinfo {author} {\bibfnamefont {J.}~\bibnamefont {Kawanaka}},\ }\bibfield  {title} {\bibinfo {title} {Velocity and acceleration freely tunable straight-line propagation light bullet},\ }\href@noop {} {\bibfield  {journal} {\bibinfo  {journal} {Scientific Reports}\ }\textbf {\bibinfo {volume} {10}},\ \bibinfo {pages} {11481} (\bibinfo {year} {2020}{\natexlab{a}})}\BibitemShut {NoStop}%
\bibitem [{\citenamefont {Li}\ and\ \citenamefont {Kawanaka}(2020{\natexlab{b}})}]{li2020optical}%
  \BibitemOpen
  \bibfield  {author} {\bibinfo {author} {\bibfnamefont {Z.}~\bibnamefont {Li}}\ and\ \bibinfo {author} {\bibfnamefont {J.}~\bibnamefont {Kawanaka}},\ }\bibfield  {title} {\bibinfo {title} {Optical wave-packet with nearly-programmable group velocities},\ }\href@noop {} {\bibfield  {journal} {\bibinfo  {journal} {Communications Physics}\ }\textbf {\bibinfo {volume} {3}},\ \bibinfo {pages} {211} (\bibinfo {year} {2020}{\natexlab{b}})}\BibitemShut {NoStop}%
\bibitem [{\citenamefont {Bhaduri}\ \emph {et~al.}(2020)\citenamefont {Bhaduri}, \citenamefont {Yessenov},\ and\ \citenamefont {Abouraddy}}]{bhaduri2020anomalous}%
  \BibitemOpen
  \bibfield  {author} {\bibinfo {author} {\bibfnamefont {B.}~\bibnamefont {Bhaduri}}, \bibinfo {author} {\bibfnamefont {M.}~\bibnamefont {Yessenov}},\ and\ \bibinfo {author} {\bibfnamefont {A.~F.}\ \bibnamefont {Abouraddy}},\ }\bibfield  {title} {\bibinfo {title} {Anomalous refraction of optical spacetime wave packets},\ }\href@noop {} {\bibfield  {journal} {\bibinfo  {journal} {Nature Photonics}\ }\textbf {\bibinfo {volume} {14}},\ \bibinfo {pages} {416} (\bibinfo {year} {2020})}\BibitemShut {NoStop}%
\bibitem [{\citenamefont {Guo}\ \emph {et~al.}(2021)\citenamefont {Guo}, \citenamefont {Xiao}, \citenamefont {Orenstein},\ and\ \citenamefont {Fan}}]{guo2021structured}%
  \BibitemOpen
  \bibfield  {author} {\bibinfo {author} {\bibfnamefont {C.}~\bibnamefont {Guo}}, \bibinfo {author} {\bibfnamefont {M.}~\bibnamefont {Xiao}}, \bibinfo {author} {\bibfnamefont {M.}~\bibnamefont {Orenstein}},\ and\ \bibinfo {author} {\bibfnamefont {S.}~\bibnamefont {Fan}},\ }\bibfield  {title} {\bibinfo {title} {Structured 3d linear space--time light bullets by nonlocal nanophotonics},\ }\href@noop {} {\bibfield  {journal} {\bibinfo  {journal} {Light: Science \& Applications}\ }\textbf {\bibinfo {volume} {10}},\ \bibinfo {pages} {160} (\bibinfo {year} {2021})}\BibitemShut {NoStop}%
\bibitem [{\citenamefont {Guo}\ and\ \citenamefont {Fan}(2021)}]{guo2021generation}%
  \BibitemOpen
  \bibfield  {author} {\bibinfo {author} {\bibfnamefont {C.}~\bibnamefont {Guo}}\ and\ \bibinfo {author} {\bibfnamefont {S.}~\bibnamefont {Fan}},\ }\bibfield  {title} {\bibinfo {title} {Generation of guided space-time wave packets using multilevel indirect photonic transitions in integrated photonics},\ }\href@noop {} {\bibfield  {journal} {\bibinfo  {journal} {Phys. Rev. Res.}\ }\textbf {\bibinfo {volume} {3}},\ \bibinfo {pages} {033161} (\bibinfo {year} {2021})}\BibitemShut {NoStop}%
\bibitem [{\citenamefont {Yessenov}\ \emph {et~al.}(2022)\citenamefont {Yessenov}, \citenamefont {Free}, \citenamefont {Chen}, \citenamefont {Johnson}, \citenamefont {Lavery}, \citenamefont {Alonso},\ and\ \citenamefont {Abouraddy}}]{yessenov2022space}%
  \BibitemOpen
  \bibfield  {author} {\bibinfo {author} {\bibfnamefont {M.}~\bibnamefont {Yessenov}}, \bibinfo {author} {\bibfnamefont {J.}~\bibnamefont {Free}}, \bibinfo {author} {\bibfnamefont {Z.}~\bibnamefont {Chen}}, \bibinfo {author} {\bibfnamefont {E.~G.}\ \bibnamefont {Johnson}}, \bibinfo {author} {\bibfnamefont {M.~P.}\ \bibnamefont {Lavery}}, \bibinfo {author} {\bibfnamefont {M.~A.}\ \bibnamefont {Alonso}},\ and\ \bibinfo {author} {\bibfnamefont {A.~F.}\ \bibnamefont {Abouraddy}},\ }\bibfield  {title} {\bibinfo {title} {Space-time wave packets localized in all dimensions},\ }\href@noop {} {\bibfield  {journal} {\bibinfo  {journal} {Nature Communications}\ }\textbf {\bibinfo {volume} {13}},\ \bibinfo {pages} {4573} (\bibinfo {year} {2022})}\BibitemShut {NoStop}%
\bibitem [{\citenamefont {Pang}\ \emph {et~al.}(2022)\citenamefont {Pang}, \citenamefont {Zou}, \citenamefont {Song}, \citenamefont {Karpov}, \citenamefont {Yessenov}, \citenamefont {Zhao}, \citenamefont {Minoofar}, \citenamefont {Zhang}, \citenamefont {Song}, \citenamefont {Zhou} \emph {et~al.}}]{pang2022synthesis}%
  \BibitemOpen
  \bibfield  {author} {\bibinfo {author} {\bibfnamefont {K.}~\bibnamefont {Pang}}, \bibinfo {author} {\bibfnamefont {K.}~\bibnamefont {Zou}}, \bibinfo {author} {\bibfnamefont {H.}~\bibnamefont {Song}}, \bibinfo {author} {\bibfnamefont {M.}~\bibnamefont {Karpov}}, \bibinfo {author} {\bibfnamefont {M.}~\bibnamefont {Yessenov}}, \bibinfo {author} {\bibfnamefont {Z.}~\bibnamefont {Zhao}}, \bibinfo {author} {\bibfnamefont {A.}~\bibnamefont {Minoofar}}, \bibinfo {author} {\bibfnamefont {R.}~\bibnamefont {Zhang}}, \bibinfo {author} {\bibfnamefont {H.}~\bibnamefont {Song}}, \bibinfo {author} {\bibfnamefont {H.}~\bibnamefont {Zhou}}, \emph {et~al.},\ }\bibfield  {title} {\bibinfo {title} {Synthesis of near-diffraction-free orbital-angular-momentum space-time wave packets having a controllable group velocity using a frequency comb},\ }\href@noop {} {\bibfield  {journal} {\bibinfo  {journal} {Opt. Express}\ }\textbf {\bibinfo {volume} {30}},\ \bibinfo {pages} {16712} (\bibinfo {year} {2022})}\BibitemShut {NoStop}%
\bibitem [{\citenamefont {Chen}\ \emph {et~al.}(2022)\citenamefont {Chen}, \citenamefont {Zhang}, \citenamefont {Liu}, \citenamefont {Meng}, \citenamefont {Dudley},\ and\ \citenamefont {Lu}}]{chen2022time}%
  \BibitemOpen
  \bibfield  {author} {\bibinfo {author} {\bibfnamefont {W.}~\bibnamefont {Chen}}, \bibinfo {author} {\bibfnamefont {W.}~\bibnamefont {Zhang}}, \bibinfo {author} {\bibfnamefont {Y.}~\bibnamefont {Liu}}, \bibinfo {author} {\bibfnamefont {F.-C.}\ \bibnamefont {Meng}}, \bibinfo {author} {\bibfnamefont {J.~M.}\ \bibnamefont {Dudley}},\ and\ \bibinfo {author} {\bibfnamefont {Y.-Q.}\ \bibnamefont {Lu}},\ }\bibfield  {title} {\bibinfo {title} {Time diffraction-free transverse orbital angular momentum beams},\ }\href@noop {} {\bibfield  {journal} {\bibinfo  {journal} {Nature Communications}\ }\textbf {\bibinfo {volume} {13}},\ \bibinfo {pages} {4021} (\bibinfo {year} {2022})}\BibitemShut {NoStop}%
\bibitem [{\citenamefont {Yessenov}\ and\ \citenamefont {Abouraddy}(2023)}]{yessenov2023relativistic}%
  \BibitemOpen
  \bibfield  {author} {\bibinfo {author} {\bibfnamefont {M.}~\bibnamefont {Yessenov}}\ and\ \bibinfo {author} {\bibfnamefont {A.~F.}\ \bibnamefont {Abouraddy}},\ }\bibfield  {title} {\bibinfo {title} {Relativistic transformations of quasi-monochromatic paraxial optical beams},\ }\href@noop {} {\bibfield  {journal} {\bibinfo  {journal} {Phys. Rev. A}\ }\textbf {\bibinfo {volume} {107}},\ \bibinfo {pages} {042221} (\bibinfo {year} {2023})}\BibitemShut {NoStop}%
\bibitem [{\citenamefont {Yessenov}\ \emph {et~al.}(2024)\citenamefont {Yessenov}, \citenamefont {Romer}, \citenamefont {Ichiji},\ and\ \citenamefont {Abouraddy}}]{yessenov2024experimental}%
  \BibitemOpen
  \bibfield  {author} {\bibinfo {author} {\bibfnamefont {M.}~\bibnamefont {Yessenov}}, \bibinfo {author} {\bibfnamefont {M.}~\bibnamefont {Romer}}, \bibinfo {author} {\bibfnamefont {N.}~\bibnamefont {Ichiji}},\ and\ \bibinfo {author} {\bibfnamefont {A.~F.}\ \bibnamefont {Abouraddy}},\ }\bibfield  {title} {\bibinfo {title} {Experimental realization of lorentz boosts of space-time wave packets},\ }\href@noop {} {\bibfield  {journal} {\bibinfo  {journal} {Phys. Rev. A}\ }\textbf {\bibinfo {volume} {109}},\ \bibinfo {pages} {013509} (\bibinfo {year} {2024})}\BibitemShut {NoStop}%
\bibitem [{\citenamefont {Gaafar}\ \emph {et~al.}(2019)\citenamefont {Gaafar}, \citenamefont {Baba}, \citenamefont {Eich},\ and\ \citenamefont {Petrov}}]{gaafar2019front}%
  \BibitemOpen
  \bibfield  {author} {\bibinfo {author} {\bibfnamefont {M.~A.}\ \bibnamefont {Gaafar}}, \bibinfo {author} {\bibfnamefont {T.}~\bibnamefont {Baba}}, \bibinfo {author} {\bibfnamefont {M.}~\bibnamefont {Eich}},\ and\ \bibinfo {author} {\bibfnamefont {A.~Y.}\ \bibnamefont {Petrov}},\ }\bibfield  {title} {\bibinfo {title} {Front-induced transitions},\ }\href@noop {} {\bibfield  {journal} {\bibinfo  {journal} {Nature Photonics}\ }\textbf {\bibinfo {volume} {13}},\ \bibinfo {pages} {737} (\bibinfo {year} {2019})}\BibitemShut {NoStop}%
\bibitem [{sup()}]{suppmat}%
  \BibitemOpen
  \href@noop {} {\bibinfo {title} {See supplemental material at [url will be inserted by publisher] for the plane wave solution at a spatiotemporal boundary, discussion of the behavior near an optical analog point of the event horizon, the transformation of spatiotemporal correlation across a spatiotemporal boundary, and a derivation of a simple model governing the light bullet trajectory, which include {Refs.} [65-66]}}\BibitemShut {NoStop}%
\bibitem [{\citenamefont {de~Sterke}(1992)}]{de1992optical}%
  \BibitemOpen
  \bibfield  {author} {\bibinfo {author} {\bibfnamefont {C.~M.}\ \bibnamefont {de~Sterke}},\ }\bibfield  {title} {\bibinfo {title} {Optical push broom},\ }\href@noop {} {\bibfield  {journal} {\bibinfo  {journal} {Opt. Lett.}\ }\textbf {\bibinfo {volume} {17}},\ \bibinfo {pages} {914} (\bibinfo {year} {1992})}\BibitemShut {NoStop}%
\end{thebibliography}%



\end{document}